\documentstyle[12pt]{article}
\begin{document}
\title{\bf\Large
The algebraic Bethe ansatz for open  vertex models }

\author{ {\bf Guang-Liang Li$^{a}$ , Kang-Jie Shi$^{b}$
} \\
{ \normalsize \it $^a$ Department of Applied Physics,
Xi'an Jiaotong University, Xi'an, 710049, P.R.China }\\
{\normalsize \it$^b$ Institute of Modern Physics, Northwest
University, Xi'an, 710069, P.R.China }\\
{\it E-mail: { \rm leegl@mail.xjtu.edu.cn}}}

\maketitle

 ABSTRACT:   We present a unified algebraic Bethe ansatz for open vertex models which are associated with  the non-exceptional
  $A^{(2)}_{2n},A^{(2)}_{2n-1},B^{(1)}_n,C^{(1)}_n,D^{(1)}_{n}$ Lie
 algebras. By  the method, we  solve these models with the trivial K
 matrix and find that our results agree with that obtained by analytical
 Bethe ansatz.  We also solve the $B^{(1)}_n,C^{(1)}_n,D^{(1)}_{n}$ models
 with some non-trivial diagonal K-matrices (one free parameter case)  by
the algebraic Bethe ansatz.

\maketitle

\vspace{10mm}


\newpage
\tableofcontents
\section{Introduction}

\noindent

The algebraic Bethe ansatz(ABA)\cite{fad}-\cite{8} has been proven
to be a more powerful mathematical method in constructing and
solving integrable models. For multi-particles models, other
methods such as the co-ordinate Bethe anzatz\cite{alcar}, become
too cumbersome, particulary for open boundary conditions. One
important feature of the ABA method is that it permits us to
present an algebraic formulation of the Bethe states. The
construction of exact eigenvectors, besides being an interesting
problem on its own, is certainly an important step in the program
of solving integrable systems. This step, however, depends much on
our ability to disentangle the Yang-Baxter algebra in terms of
appropriate commutation relations. The simplest structure of
commutation rules has been discovered in the context of six-vertex
model\cite{fad},\cite{8} and its multistate
generalization\cite{an},\cite{ano}.

However, there are many models such as
 $B_n, C_n, D_n$ vertex models whose
eigenvectors are rather complicated to be constructed.
 Fortunately, Martins and Ramos successfully generalized the ABA  and threw light
on how to construct such general muti-particle eigenvectors. They
presented a unified algebraic Bethe ansatz for a large family of
vertex models with periodic boundary and
 solved them in the way of the ABA \cite{9}-\cite{12}.

In the framework of Martins's work, recently, the method has been
applied to some nineteen-vertex like models and Hubbard-like
models models with open boundary conditions. In
refs.\cite{13}-\cite{osp12}, the eigenvectors are explicitly
constructed and the algebraic Bethe ansatz is demonstrated. The
results show that the ABA could suit some systems with higher rank
algebra symmetry.

There are many models already solved by the so  called analytical
Bethe ansatz which  was  firstly proposed by Reshetikhin for close
chains\cite{1}, and then generalized by Nepomechie and his
collaborators into the quantum-algebra-invariant open chains
\cite{2}. Later,  a set of open vertex models associated with
non-exceptional Lie algebras such as $A^{(2)}_{2n}$,
$A^{(2)}_{2n-1}$, $B^{(1)}_n$, $C^{(1)}_n$, $D^{(1)}_{n}$ vertex
models\cite{a2n2} were solved by the analytical Bethe ansatz.
Compared with  the analytical Bethe ansatz, the ABA can provide us
more rigorous results. So it is well worth reconsidering these
models by the ABA.

 In this paper we will consider the
 $A^{(2)}_{2n}$, $A^{(2)}_{2n-1}$, $B^{(1)}_n$, $C^{(1)}_n$,
$D^{(1)}_{n}$  vertex models with the trivial K matrix and
$B^{(1)}_n,C^{(1)}_n,D^{(1)}_{n}$ models with some
 non-trivial  diagonal K-matrices(one free parameter case). These
models with period boundary conditions have been solved by
Reshetikhin in terms of analytical Bethe ansatz and by Lima-santos
with the help of the ABA\cite{lima}, respectively. Their
quantum-algebra-invariant open chains (corresponding to the
trivial diagonal reflecting K matrix) have already been considered
by Nepomechie et al.\cite{a2n2}. Here, we will present a unified
ABA for these models. We find that the fundamental commutation
relations rules have a common form in terms of the corresponding
Boltzmann weights. As a consequence, the derivations of the
eigenvectors, the eigenvalues and the associated Bethe ansatz
equations also have a quite general character for these vertex
models.  Our results agree with that obtained by the analytical
Bethe ansatz.

We organize our  paper   as following. In section 2 we demonstrate
the ABA  for the open $A^{(2)}_{2n}$, $A^{(2)}_{2n-1}$,
$B^{(1)}_n$, $C^{(1)}_n$, $D^{(1)}_{n}$  vertex models and present
the solutions to these models with the trivial K matrix. The
results for $B^{(1)}_n,C^{(1)}_n,D^{(1)}_{n}$ models with
 one free parameter diagonal K-matrices are also obtained. A brief
summary and discussion about our  results are included in section
3. Some coefficients and detail derivations  are given as the
Appendices.


\section{The algebraic  Bethe ansatz}

\subsection{The R matrices and K matrices }
 The R matrices for  the
$A^{(2)}_{2n},A^{(2)}_{2n-1},B^{(1)}_n,C^{(1)}_n,D^{(1)}_{n}$
models\cite{vv},\cite{mj} can be expressed as \cite{a2n2}
 \begin{eqnarray}
   \hspace{-2mm}R^{(n)}(u) &=&a_n(u)\sum_{i\ne \bar{i}}E_{ii}\otimes
  E_{ii}+b_n(u)\sum_{i\ne j,\bar{j}}E_{ii}\otimes E_{jj}
   +\Big(\sum_{i<\bar{i}}c_n(u,i)+\sum_{i>\bar{i}}\bar{c}_n(u,i)\Big)
   E_{i\bar{i}}\otimes E_{\bar{i}i} \nonumber\\
   & &+
   \Big(\sum_{i<j,j\ne\bar{i}}d_n(u,i,j)+\sum_{i>j,j\ne\bar{i}}\bar{d}_n(u,i,j)\Big)
   E_{i{j}}\otimes E_{\bar{i}\bar{j}}
   +e_n(u)\sum_{i\ne \bar{i}}E_{ii}\otimes E_{\bar{i}\hspace{0.2mm} \bar{i}}
   \nonumber\\
   & &+\delta_{i\bar{i}}f_n(u)E_{i{i}}\otimes E_{\bar{i}\hspace{0.2mm}\bar{i}}+
   \Big(g_n(u)\sum_{i<j,j\ne\bar{i}}+\bar{g}_n(u)\sum_{i>j,j
   \ne\bar{i}}\Big)E_{ij}\otimes E_{{j}i} .
 \label{rm}
 \end{eqnarray}
The  summations run over 1 to $q$, $i+\bar{i}=q+1$, $q=2n+1$ for
$A^{(2)}_{2n} , B_n^{(1)}$  and  $q=2n$ for
$A^{(2)}_{2n-1},C_n^{(1)},D_n^{(1)}$,
$(E_{ij})_{kl}=\delta_{ik}\delta_{jl}$. The R matrices satisfy the
following properties
\begin{eqnarray}
regularity&:&R^{(n)}_{12}(0)=\rho_n(0)^{\frac{1}{2}}{\cal P}_{12},\nonumber\\
unitarity&:&R^{(n)}_{12}(u)R^{(n)}_{21}(-u)=\rho_n(u),\nonumber\\
PT-symmetry&:&{\cal P}_{12}R^{(n)}_{12}(u){\cal P}_{12}=[R^{(n)}_{12}]^{t_1t_2}(u),\nonumber\\
crossing-unitarity&:&M_1{(n)}R^{(n)}_{12}(u)^{t_2}{M_1{(n)}}^{-1}R^{(n)}_{12}(-u-2\vartheta_n)^{t_1}=
\rho_n(u+\vartheta_n).\nonumber
\end{eqnarray}
 Where $\rho_n(u)=a_n(u)a_n(-u)$, ${\cal
P}={\cal P}^{ik}_{jl}E_{ij}\otimes E_{kl}$, ${\cal
P}^{ik}_{jl}=\delta_{il}\delta_{jk}$, $
\vartheta_n=-\sqrt{-1}\pi-2\kappa\eta$ for
$A^{(2)}_{2n},A^{(2)}_{2n-1}$ and $ \vartheta_n=-2\kappa\eta$ for
$B_n^{(1)},C_n^{(1)},D_n^{(1)}$,  $\kappa=2n+1,2n,2n-1,2n+2,2n-2$
for $A^{(2)}_{2n},A^{(2)}_{2n-1},B_n^{(1)},C_n^{(1)},D_n^{(1)}$,
respectively. $t_i$ denotes the transposition in $i$-th space,
$M_1{(n)}=M{(n)}\otimes 1$ and
\begin{equation}
M_{ij}(n)=\left\{\begin{array}{ll}
\delta_{ij}e^{4(n+1-\tilde{i})\eta} & \mbox{for}
A^{(2)}_{2n},\hspace{2mm}B_n^{(1)}\\
\delta_{ij}e^{4(n+1/2-\tilde{i})\eta} & \mbox{for}
A^{(2)}_{2n-1},\hspace{2mm}C_n^{(1)},\hspace{2mm}D_n^{(1)}
\end{array}\right.
\end{equation}
with
\begin{eqnarray}
&&\qquad\tilde{i}=\left\{\begin{array}{ll}
i+\frac{1}{2},& 1\le i< \frac{q+1}{2}\\
i,& i=\frac{q+1}{2}\\
i-\frac{1}{2},& \frac{q+1}{2}< i\le q, \hspace{1cm}\mbox{for
$A^{(2)}_{2n}, B_n^{(1)}, D_n^{(1)}$}
\end{array}\right.\\
&&\qquad\tilde{i}=\left\{\begin{array}{ll}
i-\frac{1}{2},& 1\le i\le n\\
i+\frac{1}{2},& n+1\le i\le 2n. \hspace{1cm}\mbox{for
$A^{(2)}_{2n-1}, C_n^{(1)}$}
\end{array}\right.
\end{eqnarray}
Here one should notice that $A^{(2)}_{2n-1}$ R matrix we used is
$U_q(C_n)$ invariant, instead of  $U_q(D_n)$ invariant  in
refs.\cite{mj},\cite{A2n},\cite{A2n1}. These R matrices satisfy
the Yang-Baxter equation(YBE)\cite{6}
\begin{eqnarray}
R^{(n)}_{12}(u-v)R^{(n)}_{13}(u)R^{(n)}_{23}(v)=R^{(n)}_{23}(v)R^{(n)}_{13}(u)R^{(n)}_{12}(u-v),
\label{ybe}
\end{eqnarray}
$R^{(n)}_{12}(u)=R^{(n)}(u)\otimes 1, R^{(n)}_{23}(u)=1\otimes
R^{(n)}(u)$ etc., $R^{(n)}_{21}={\cal P}_{12}R^{(n)}_{12}{\cal
P}_{12}$. For a $N\times N$ square lattice, if we can find the $K$
matrices satisfying the so called reflection equations
\begin{eqnarray}
 && R^{(n)}_{12}(u-v)\stackrel{1}{K}_{-}(u)R^{(n)}_{21}(u+v) \stackrel{2}{K}_{-}(v)=
 \stackrel{2}{K}_{-}(v)R^{(n)}_{12}(u+v)\stackrel{1}{K}_{-}(u)R^{(n)}_{21}(u-v),
 \label{r1}\\
 & &R^{(n)}_{12}(-u+v)\stackrel{1}{K}^{t_1}_{+}(u)\stackrel{1}{M^{-1}}(n)R^{(n)}_{21}
 (-u-v-2\xi_n)\stackrel{1}{M}(n)\stackrel{2}{K}^{t_2}_{+}(v)\nonumber\\
&&=\stackrel{2}{K}^{t_2}_{+}(v)\stackrel{1}{M}(n)R^{(n)}_{12}(-u-v-2\xi_n)\stackrel{1}{M^{-1}}(n)
\stackrel{1}{K}^{t_1}_{+}(u)R^{(n)}_{21}(-u+v),
 \label{r2}
 \end{eqnarray}
where $\stackrel{1}{K}_{\pm}(u)=K_{\pm}(u)\otimes
1,\stackrel{2}{K}_{\pm}(u)=1\otimes K_{\pm}(u)$, then the transfer
matrix  defined as
\begin{equation}
t(u)={\rm tr}K_{+}(u)U(u) \label{tru}
\end{equation}
constitutes an one-parameter commutative family, i.e.
$[t(u),t(v)]=0$. Here
\begin{eqnarray}
U(u)&=&T(u)K_{-}(u)T^{-1}(-u),\label{Uform}\\
T(u)&=&R^{(n)}_{01}(u)R^{(n)}_{02}\cdots
R^{(n)}_{0N}(u).\label{Tform}
\end{eqnarray}
As indicated by Sklynanin \cite{8},  the transfer matrix is
related to the Hamiltonian of quantum chain with the
nearest-neighbour interaction and boundary terms through the
following relation
\begin{equation}
t'(0)=c_{\zeta}[2H trK_{+}(0)+ trK'_{+}(0)]
\end{equation}
and the corresponding integrable open chain Hamiltonian takes the
form
\begin{equation}
H=\sum^{N-1}_{k=1}H_{k,k+1}
+\frac{1}{2c_{\zeta}}\stackrel{1}{K'}_{-}(0)+\frac{{\rm
tr}\stackrel{0}{K_{+}}(0)H_{N,0}}{{\rm tr}K_{+}(0)} \label{hh}
\end{equation}
with $H_{k,k+1}={\cal P}_{k,k+1}R_{kk+1}'(u)|_{u=0}$. $c_{\zeta}$
is a constant defined by $K_{-}(0)=c_{\zeta}\cdot id$.

 From
Eq.(\ref{r1}) and Eq.(\ref{r2}), we can see that, given a solution
$K_{-}(u)$ of Eq.(\ref{r1}), the matrix
\begin{equation}
K_{+}(u)=K_{-}(-u-\vartheta)^tM{(n)} \label{kk}
\end{equation}
satisfies Eq.(\ref{r2}). The general solutions to Eq(\ref{r1}) for
$A^{(2)}_{2n},A^{(2)}_{2n-1},B^{(1)}_n,C^{(1)}_n,D^{(1)}_{n}$
models have been obtained in refs. \cite{A2n},\cite{A2n1}. For
Eq.(\ref{r1}), there are three kinds of diagonal K matrices:  the
trivial  K matrix,  K matrices without free parameter and
 K matrices with one free parameter.
We denote these K matrices as $K^{(1)}_{-}(n)$,
$K^{(2)}_{-}({u},n,p_{-})$ and $K^{(3)}_{-}({u},n,{\zeta}_{-})$,
respectively, where $p_{-}$ are integer number, ${\zeta}_{-}$ are
free parameters. $K^{(1)}_{-}(n)$ is a $q\times q $ identity
matrix and can be written as
\begin{equation}
K^{(1)}_{-}(n)=1. \label{iden}
\end{equation}
For  the $A^{(2)}_{2n}$ model with the non-trivial diagonal
K-matrices have been solved in ref.\cite{osp} and the
$A^{(2)}_{2n-1}$ R matrix we used is different from that in
refs.\cite{A2n},\cite{A2n1}, therefore, we do not consider this
two models with non-trivial diagonal K-matrices in this paper. We
only present the non-trivial diagonal K-matrices obtained in
refs.\cite{A2n},\cite{A2n1} for the
$B^{(1)}_n,C^{(1)}_n,D^{(1)}_{n}$ models as below,
\begin{eqnarray}
K^{(2)}_{-}(u,n,p_-)_{i} & =&\left\{\begin{array}{l}
           e^{-u}\sinh((2p_--n-n_c)\eta-\frac{u}{2}),
                 \quad(1\le i\le p_-)\\[2mm]
           \sinh((2p_--n-n_c)\eta+\frac{u}{2}),
                \quad(p_-+1\le i\le q-p_-)\\[2mm]
           e^{u}\sinh((2p_--n-n_c)\eta-\frac{u}{2}).
                 \quad(q+1-p_-\le i\le q)\end{array}\right.
                 \label{kn21bn}\\[5mm]
K^{(2)}_{-}(u,n,p_-)_{i} & =&\left\{\begin{array}{l}
           e^{-u}\cosh((2p_--n-n_c)\eta-\frac{u}{2}),
                 \quad(1\le i\le p_-)\\[2mm]
           \cosh((2p_--n-n_c)\eta+\frac{u}{2}),
                \quad(p_-+1\le i\le q-p_-)\\[2mm]
           e^{u}\cosh((2p_--n-n_c)\eta-\frac{u}{2}).
                 \quad(q+1-p_-\le i\le q)\end{array}\right.\label{kn22bn}
\end{eqnarray}
where $n_c=\frac{1}{2}$ for $B_n^{(1)}$ and $n_c={2}$ for
$C_n^{(1)},D_n^{(1)}$, $p_-$ ranges in $[2,n]$ and $[2,n-1]$ for
$B_n^{(1)},C_n^{(1)}$ and $D_n^{(1)}$, respectively. For
$C_n^{(1)},D_n^{(1)}$, Eq.(\ref{kn21bn}) do not hold when
$2p_-=n+2$. The diagonal K matrices with one free parameter for
$B_n^{(1)}$ and $C_n^{(1)}$ are
\begin{eqnarray}
K^{(3)}_{-}(u,n,\zeta_-)_{i} & =&\left\{\begin{array}{l}
           e^{-u}\sinh(\zeta_-+\frac{u}{2})\sinh(\zeta_--(2n-3)\eta+\frac{u}{2}),
                 \quad(i=1)\\[2mm]
           \sinh(\zeta_--\frac{u}{2})\sinh(\zeta_--(2n-3)\eta+\frac{u}{2})
                \quad(2\le i\le 2n)\\[2mm]
           e^{u}\sinh(\zeta_--\frac{u}{2})\sinh(\zeta_--(2n-3)\eta-\frac{u}{2})
                 \quad(i=2n+1)\end{array}\right.
                 \label{kn3bn}
\end{eqnarray}
and
\begin{eqnarray}
K^{(3)}_{-}(u,n,\zeta_-)_{i} & =&\left\{\begin{array}{l}
           e^{-\frac{u}{2}}\sinh(\zeta_-+\frac{u}{2}),
                 \quad(1\le i\le n)\\[2mm]
           e^{\frac{u}{2}}\sinh(\zeta_--\frac{u}{2}),
                 \quad(n+1\le i\le 2n)\end{array}\right.
                 \label{kn3cn}
\end{eqnarray}
respectively.  For  $D_n^{(1)}$, we can get
\begin{eqnarray}
K^{(3)}_{-}(u,n,\zeta_-)_{i} & =&\left\{\begin{array}{l}
           e^{-u}\sinh(\zeta_-+\frac{u}{2})\sinh(\zeta_--(2n-4)\eta+\frac{u}{2}),
                 \quad(i=1)\\[2mm]
           \sinh(\zeta_--\frac{u}{2})\sinh(\zeta_--(2n-4)\eta+\frac{u}{2})
                \quad(2\le i\le 2n)\\[2mm]
           e^{u}\sinh(\zeta_--\frac{u}{2})\sinh(\zeta_--(2n-4)\eta-\frac{u}{2}),
                 \quad(i=2n+1)\end{array}\right.
                 \label{kn31}\\[5mm]
K^{(3)}_{-}(u,n,\zeta_-)_{i} & =&\left\{\begin{array}{l}
           e^{-\frac{u}{2}}\sinh(\zeta_-+\frac{u}{2}),
                 \quad(1\le i\le n)\\[2mm]
           e^{\frac{u}{2}}\sinh(\zeta_--\frac{u}{2}),
                 \quad(n+1\le i\le 2n)\end{array}\right.
                 \label{kn32}\\[5mm]
K^{(3)}_{-}(u,n,\zeta_-)_{i} & =&\left\{\begin{array}{l}
           e^{-\frac{u}{2}}\sinh(\zeta_-+\frac{u}{2}),
                 \quad(i=1,2,\cdots,n-1,n+1)\\[2mm]
           e^{\frac{u}{2}}\sinh(\zeta_--\frac{u}{2}),
                 \quad(i=n,n+2,\cdots,2n)\end{array}\right.
                 \label{kn33}\\[5mm]
K^{(3)}_{-}(u,n,\zeta_-)_{i} & =&\left\{\begin{array}{l}
           e^{-\frac{u}{2}}\sinh(\zeta_-+\frac{u}{2}),
                 \quad(i=1)\\[2mm]
           e^{\frac{u}{2}}\sinh(\zeta_--\frac{u}{2}),
                 \quad(i=2,\cdots,n-1,n)\\[2mm]
           e^{\frac{3u}{2}}\sinh(\zeta_-+\frac{u}{2}),
                 \quad(i=n+1,n+2,\cdots,2n-1)\\[2mm]
           e^{\frac{5u}{2}}\sinh(\zeta_--\frac{u}{2}),
                 \quad(i=2n)\end{array}\right.
                 \label{kn34}\\[5mm]
K^{(3)}_{-}(u,n,\zeta_-)_{i} & =&\left\{\begin{array}{l}
           e^{-\frac{u}{2}}\sinh(\zeta_-+\frac{u}{2}),
                 \quad(i=1)\\[2mm]
           e^{\frac{u}{2}}\sinh(\zeta_--\frac{u}{2}),
                 \quad(i=2,\cdots,n-1,n+1)\\[2mm]
           e^{\frac{3u}{2}}\sinh(\zeta_-+\frac{u}{2}),
                 \quad(i=n,n+2,\cdots,2n-1)\\[2mm]
           e^{\frac{5u}{2}}\sinh(\zeta_--\frac{u}{2}),
                 \quad(i=2n)\end{array}\right.
                 \label{kn35}\\[5mm]
K^{(3)}_{-}(u,n,\zeta_-)_{i} & =&\left\{\begin{array}{l}
           e^{-u}\sinh(\zeta_-+\frac{u}{2})\sinh(\zeta_-+(2n-4)\eta-\frac{u}{2}),
                 \quad(i=1,2,\cdots, n-1)\\[2mm]
           \sinh(\zeta_-+\frac{u}{2})\sinh(\zeta_-+(2n-4)\eta+\frac{u}{2}),
                 \quad(i=n+1)\\[2mm]
           \sinh(\zeta_--\frac{u}{2})\sinh(\zeta_-+(2n-4)\eta-\frac{u}{2}),
                 \quad(i=n)\\[2mm]
           e^{u}\sinh(\zeta_-+\frac{u}{2})\sinh(\zeta_-+(2n-4)\eta-\frac{u}{2}),
                 \quad(i=n+2,\cdots, 2n)\end{array}\right.
                 \label{kn36}\\[5mm]
K^{(3)}_{-}(u,n,\zeta_-)_{i} & =&\left\{\begin{array}{l}
           e^{-u}\sinh(\zeta_-+\frac{u}{2})\sinh(\zeta_-+(2n-4)\eta-\frac{u}{2}),
                 \quad(i=1,2,\cdots, n-1)\\[2mm]
           \sinh(\zeta_-+\frac{u}{2})\sinh(\zeta_-+(2n-4)\eta+\frac{u}{2}),
                 \quad(i=n)\\[2mm]
           \sinh(\zeta_--\frac{u}{2})\sinh(\zeta_-+(2n-4)\eta-\frac{u}{2}),
                 \quad(i=n+1)\\[2mm]
           e^{u}\sinh(\zeta_-+\frac{u}{2})\sinh(\zeta_-+(2n-4)\eta-\frac{u}{2}).
                 \quad(i=n+2,\cdots, 2n)\end{array}\right.
                 \label{kn37}
\end{eqnarray}
Correspondingly, from  Eq.(\ref{kk}) and
Eqs.(\ref{iden}-\ref{kn37}), we can obtain three kinds of diagonal
$K_+$ matrices: $K^{(1)}_{+}(n)$, $K^{(2)}_{+}({u},n,p_{+})$ and
$K^{(3)}_{+}({u},n,{\zeta}_{+})$ for Eq.(\ref{r2}). Where $p_{+}$
are integer number, ${\zeta}_{+}$ are free parameters.

\subsection{The vacuum state and commutation relations}
Firstly,  we write the double-monodromy matrix (\ref{Uform}) as
\begin{equation}
U(u)=\left(\begin{array}{cccccc}
 A(u)& B_1(u)& B_2(u)& \cdots & B_{2n-1}&F(u)\\
D_1(u)& A_{11}(u)& A_{12}(u)&\cdots &A_{1q-2}(u) &E_1(u)\\
D_2(u)& A_{21}(u)& A_{22}(u)&\cdots &A_{2q-2}(u) &E_2(u)\\
\vdots&\vdots&      \vdots&  \vdots& \vdots&\vdots \\
D_{q-2}(u)& A_{q-21}(u)& A_{q-22}(u)&\cdots &A_{q-2q-2}(u) &E_{q-2}(u)\\
 G(u)& C_1(u)& C_2(u)&\cdots &C_{q-2}(u) & A_2(u)
\end{array}\right). \label{uu}
\end{equation}
With the help of Eqs.(\ref{ybe},\ref{r1}), we can prove that
$U(u)$ in Eq.(\ref{uu})  satisfy the following equation
\begin{equation}
 R^{(n)}_{12}(u-v)\stackrel{1}U(u)R^{(n)}_{21}(u+v) \stackrel{2}U(v)=
 \stackrel{2}U(v)R^{(n)}_{12}(u+v)\stackrel{1}U(u)R^{(n)}_{21}(u-v).
 \label{dbmzj}
 \end{equation}
Now we Introduce the vacuum state,
\begin{equation}
|0\rangle=\prod^{\otimes N}(1,0,\cdots,0)^t, \label{zkt}
\end{equation}
\noindent where $(1,0,\cdots,0)$ is a $1\times q$ matrix, $t$
denotes the transposition.

Applying  the double-row monodromy matrix eq.(\ref{uu}) on the
vacuum state eq.(\ref{zkt}), we can find
\begin{eqnarray}
& &D_a(u)|0\rangle=0,\hspace{4mm}C_a(u)|0\rangle=0,
\hspace{4mm}G(u)|0\rangle=0,\nonumber \\
& &B_a(u)|0\rangle\ne 0, \hspace{4mm}E_a(u)|0\rangle\ne 0,
\hspace{4mm} F(u)|0\rangle\ne 0,\nonumber \\
& &A_{aa}(u)|0\rangle\ne 0,\hspace{3mm}A_{ab}(u)|0\rangle= 0\ \
(a\ne b
),\nonumber\\
& &A(u)|0\rangle\ne 0,\hspace{5mm}A_{2}(u)|0\rangle\ne
0.\hspace{4mm} (a=1,2,\cdots,q-2)\label{avc}
\end{eqnarray}
From Eq.(\ref{avc}), we can see that $D_a,C_a$  and $B_a, E_a, F$
play the role of annihilation operators and creation operators on
the vacuum state, respectively. The $A, A_{aa}, A_2$ are diagonal
operators on the vacuum state. Considering the definition of
$U(u)$ Eq.(\ref{Uform}),
 we have

\begin{eqnarray}
A(u)|0\rangle&=& T(u)_{11}K_{-}(u)_1T^{-1}(-u)_{11}|0\rangle, \label{aa1} \\
A_{aa}(u)|0\rangle&=&
T(u)_{a+11}K_{-}(u)_1T^{-1}(-u)_{1a+1}|0\rangle\nonumber\\
&&+T(u)_{a+1a+1}K_{-}(u)_{a+1}T^{-1}(-u)_{a+1a+1}|0\rangle, \label{dd1} \\
A_{2}(u)|0\rangle&=&\sum^{q}_{i=1}
T(u)_{qi}K_{-}(u)_iT^{-1}(-u)_{iq}|0\rangle. \label{dd2}
\end{eqnarray}

In above equations, the first term of eq.(\ref{dd1}) and the
previous $q-1$ terms of eq.(\ref{dd2}) can not be calculated
directly but it can be worked out by using the following method.
Taking $v=-u$ in the Yang-Baxter equation, we can get

\begin{eqnarray}
T_2^{-1}(-u)R_{12}(2u)T_1(u)=T_1(u)R_{12}(2u)T^{-1}_2(-u).
\label{dd3}
\end{eqnarray}

\noindent Taking special indices in this relation and applying
both sides of this relation to the vacuum state, we find:

\begin{eqnarray}
& & T(u)_{a+11}T^{-1}(-u)_{1a+1}|0\rangle=\tilde{f}_1(u)
\left(T^{-1}(-u)_{11}T(u)_{11}-T(u)_{a+1a+1}T^{-1}(-u)_{a+1a+1}\right)|0\rangle, \nonumber \\
& & T(u)_{qi}T^{-1}(-u)_{iq}|0\rangle=M^{(n)}_{ii}\tilde{f}_2(u)
\left(T^{-1}(-u)_{ii}T(u)_{ii}-T(u)_{qq}T^{-1}(-u)_{qq}\right)|0\rangle,(i\ne 1,q) \nonumber \\
& &
T(u)_{q1}T^{-1}(-u)_{1q}|0\rangle=\left(\tilde{f}_3(u)T^{-1}(-u)_{11}T(u)_{11}-\tilde{f}_1(u)\tilde{f}_2(u)
\sum_{i=2}^{q-1}M^{(n)}_{ii}T(u)_{ii}T^{-1}(-u)_{ii}\right.\nonumber \\
& &
\hspace{4cm}\left.+(\tilde{f}_1(u)\tilde{f}_2(u)\sum_{i=2}^{q-1}M^{(n)}_{ii}-\tilde{f}_3(u))
T^{-1}(-u)_{qq}T(u)_{qq}\right)|0\rangle,\label{gfcb}
\end{eqnarray}
where
\begin{eqnarray}
& & \tilde{f}_1(u)=\frac{\bar{g}_n(2u)}{a_n(2u)},
\qquad\tilde{f}_3(u)=\frac{\bar{c}_{n}(2u,q)}{a_n(2u)},\nonumber\\
&&\tilde{f}_2(u)=\frac{\bar{c}_{n}(2u,q){g}_n(2u)-\bar{g}_n(2u){a}_n(2u)}
{{g}_n(2u)\bar{g}_n(2u)\sum_{i=2}^{q-1}M^{(n)}_{ii}-{a}_n(2u)\sum_{i=2}^{q-1}M^{(n)}_{ii}{R^{(n)}(2u)}^{ij}_{ji}}\nonumber\\
\end{eqnarray}
with arbitrary $j\in[2,q-1]$.  Simple calculations show that
\begin{equation}
\tilde{f}_2(u)=\left\{\begin{array}{ll}
\displaystyle-\frac{e^{u-4\eta}\sinh(2\eta)}{\sinh(u-2(\kappa-1)\eta)},&
\mbox{for
$A^{(2)}_{2n}, C^{(1)}_n$}\\
\displaystyle-\frac{e^{u}\sinh(2\eta)}{\sinh(u-2(\kappa-1)\eta)}.&
\mbox{for $A^{(2)}_{2n-1}, B^{(1)}_n, D^{(1)}_n$}
\end{array}\right.
\end{equation}
 Introducing two new operators
\begin{eqnarray}
\tilde{A}_{ab}(u)&=&A_{ab}(u)-\tilde{f}_1(u)A(u)\delta_{ab},\label{d1}\\
\tilde{A}_2(u)&=&A_2(u)-\tilde{f}_3(u)A(u)-\tilde{f}_2(u)\sum_{a=1}^{q-2}M^{(n-1)}_{aa}\tilde{A}_{aa}(u),\label{d2}
\end{eqnarray}
here we should notice that $M^{(n-1)}_{aa}=M^{(n)}_{a+1a+1}$, then
we have
\begin{eqnarray}
&&A(u)|0\rangle=K_-(u)_{1}{[a_n(u)]^{2N}}{\rho_n(u)^{-N}}
|0\rangle=\omega_1 (u)|0\rangle,\label{Avac}\\[3mm]
&&\tilde{A}_{aa}(u)|0\rangle=(K_-(u)_{a+1}-\tilde{f}_1(u)
K_-(u)_{1}){[b_n(u)]^{2N}}{\rho_n(u)^{-N}}|0\rangle
=k^-({u})_a\omega(u)|0\rangle,\label{A1vac}\\[3mm]
&&\tilde{A}_2(u)|0\rangle=\Big\{K_-(u)_{q}-\tilde{f}_2(u)
\sum_{a=1}^{q-2}M^{(n-1)}_{aa}\Big(K_-(u)_{a+1}
-\tilde{f}_1(u)K_-(u)_{1}\Big)\nonumber\\
&&\hspace{4cm}-\tilde{f}_3(u)K_-(u)_{1}\Big\}
{[e_n(u)]^{2N}}{\rho_n(u)^{-N}}|0\rangle=\omega_{q}(u)|0\rangle.\label{A3vac}
\end{eqnarray}
\noindent In terms of new operators, the transfer matrix
(\ref{tru}) can be rewritten as
\begin{eqnarray}
&&t(u)={w}_1(u)A(u)+\sum_{a=1}^{q-2}{w}(u)k^+_{a}(u)\tilde{A}_{aa}(u)
+{w}_{q}(u)\tilde{A}_2(u)\label{tru1}
\end{eqnarray}
with
\begin{eqnarray}
&&{w}_1(u)=K_+(u)_1+\tilde{f}_3(u)K_+(u)_{q}+\tilde{f}_1(u)
\sum_{a=1}^{q-2}K_+(u)_{a+1},\nonumber\\
&&{w}(u)k^+_{a}(u)=K_+(u)_{a+1}+M^{(n-1)}_{aa}\tilde{f}_2(u)K_+(u)_{q},\hspace{4mm}
{w}_{q}(u)=K_+(u)_{q}.\label{A4vac}
\end{eqnarray}
Where the nested $k^{\mp}({u})$ matrices take the forms
$K^{(1)}_{\mp}(n-1)$, $K^{(2)}_{\mp}(\tilde{u},n-1,p_{\mp}-1)$ or
$K^{(3)}_{\mp}(\tilde{u},n-1,\tilde{\zeta}_{\mp})$ depending on
the choice of boundary, $\tilde{u}=u-2\eta$,
$\tilde{\zeta}_{\mp}={\zeta}_{\mp}\pm \eta$ or ${\zeta}_{\mp}+
\eta.$ For examples, $k^{\mp}({u})$ take
$K^{(2)}_{\mp}(\tilde{u},n-1,p_{\mp}-1)$ and
$K^{(3)}_{\mp}(\tilde{u},n-1,\tilde{\zeta}_{\mp})$ for
Eqs.(\ref{kn21bn},\ref{kn22bn}) and
Eqs.(\ref{kn3cn},\ref{kn32}-\ref{kn37}), respectively, while it
take $K^{(1)}_{\mp}(n-1)$ for Eqs.(\ref{kn3bn},\ref{kn31}).

In order to construct the general m-particle state, we need to
find the commutation relations between the creation,
 diagonal and annihilation fields. Rewriting the equation (\ref{dbmzj}) in component form
\begin{eqnarray}
&&R(u_-)^{a_1a_2}_{c_1c_2}U(u)^{c_1}_{d_1}R(u_+)^{c_2d_1}_{d_2b_1}
U(v)^{d_2}_{b_2}\nonumber \\[4mm]
&&=U(v)^{a_2}_{c_2}R(u_+)^{a_1c_2}_{c_1d_2}U(u)^{c_1}_{d_1}
R(u_-)^{d_2d_1}_{b_2b_1}, \label{3ufc}
\end{eqnarray}
\noindent  where the repeated indices  sum over 1 to $q$,
$u_-=u-v$, $u_+=u+v$. In the following formulaes, the repeated
indices will mean summing over 1 to $q-2$ unless there are some
special notifications.
 Taking some components of Eq.(\ref{3ufc}), we can obtain the
following fundamental commutation relations for those models,
\begin{eqnarray}
 & & B_a(u)B_b(v)+\delta_{\bar{a}b} g_1(u,v,a)F(u)A(v)
 + g_2(u,v,a)F(u)\tilde{A}_{\bar{a}b}(v)\nonumber\\
& &=\hat{r}(u_-)^{dc}_{ba} [B_d(v)B_c(u)+\delta_{\bar{d}c}
g_1(v,u,d)F(v)A(u)
 + g_2(v,u,d)F(v)\tilde{A}_{\bar{d}c}(u)],
\label{b1b1}
\end{eqnarray}
\begin{eqnarray}
A(u)B_a(v)&=&a^1_1(u,v)B_a(v)A(u)+a^1_2(u,v)B_a(u)A(v)+a^1_3(u,v)B_d(u)\tilde{A}_{da}(v)\nonumber\\
 & &+ a^1_4(u,v,\bar{a})F(u)D_{\bar{a}}(v)+a^1_5(u,v)F(u)C_a(v)+ a^1_6(u,v,\bar{a})F(v)D_{\bar{a}}(u),\label{ab1}\\[4mm]
\tilde{A}_{ab}(u)B_c(v)&=&{\tilde{r}(u_+)^{ae}_{dg}\bar{r}(u_-)^{gf}_{cb}}B_e(v)\tilde{A}_{df}(u)
+R^{A}_1(u,v)^{af}_{cb}B_f(u)A(v)\nonumber\\
& & +R^{A}_2(u,v)^{af}_{db}B_f(u)\tilde{A}_{dc}(v)
 +\delta_{\bar{b}c} R^{A}_3(u,v,\bar{b})E_a(u)A(v)\nonumber\\
 & & + R^{A}_4(u,v,\bar{b})E_a(u)\tilde{A}_{\bar{b}c}(v)+  R^{A}_5(u,v)^{af}_{cb}F(u)D_{\bar{f}}(v)\nonumber\\
 & &+\delta_{ab}R^{A}_6(u,v)F(u)C_c(v)+ R^{A}_7(u,v)^{af}_{cb}F(v)D_{\bar{f}}(u)\nonumber\\
 & & +R^{A}_8(u,v)^{af}_{cb}F(v)C_f(u),\label{d1b1}\\[4mm]
\tilde{A}_2(u)B_a(v)&=&a^3_1(u,v)B_a(v)\tilde{A}_2(u)+a^3_2(u,v)B_a(u)A(v)+a^3_3(u,v)B_d(u)\tilde{A}_{da}(v)\nonumber\\
 & &+ a^3_4(u,v,{a})E_{\bar{a}}(u)A(v)+ a^3_5(u,v,\bar{d})E_d(u)\tilde{A}_{\bar{d}a}(v)
 +  a^3_6(u,v,\bar{a})F(u)D_{\bar{a}}(v)\nonumber\\
 & &+a^3_7(u,v)F(u)C_a(v)+
 a^3_8(u,v,\bar{a})F(v)D_{\bar{a}}(u)+a^3_{9}(u,v)F(v)C_a(u),\label{d2b1}
\end{eqnarray}
\begin{eqnarray}
A(u)F(v)&=& b^1_1(u,v)F(v)A(u)+b^1_2(u,v)F(u)A(v)+b^1_3(u,v,d)F(u)\tilde{A}_{dd}(v)\nonumber\\
 & &+b^1_4(u,v)F(u)\tilde{A}_2(v)+ b^1_5(u,v,d)B_{\bar{d}}(u)B_d(v)+b^1_6(u,v)B_d(u)E_d(v),\\[4mm]
\tilde{A}_{ab}(u)F(v)&=&b^2_1(u,v)F(v)\tilde{A}_{ab}(u)+\delta_{ab}b^2_2(u,v)F(u)A(v)
+R^{F}_1(u,v)^{dc}_{b\bar{a}}F(u)\tilde{A}_{\bar{d}c}(v)\nonumber\\
& &+\delta_{ab}b^2_3(u,v)F(u)\tilde{A}_2(v)+
R^F_2(u,v)^{dc}_{b\bar{a}}B_d(u)B_c(v)
+R^F_3(u,v)^{ac}_{db}B_c(u)E_d(v)\nonumber\\
 & &+b^2_4(u,v)E_a(u)B_b(v)+ b^2_5(u,v,\bar{b})E_a(u)E_{\bar{b}}(v),\\[4mm]
\tilde{A}_2(u)F(v)&=&b^3_1(u,v)F(v)\tilde{A}_2(u)+b^3_2(u,v)F(u)A(v)+b^3_3(u,v,d)F(u)\tilde{A}_{dd}(v)\nonumber\\
& &+b^3_4(u,v)F(u)\tilde{A}_2(v)+ b^3_5(u,v,\bar{d})B_{\bar{d}}(u)B_d(v)+b^3_6(u,v)B_d(u)E_d(v)\nonumber\\
& &+b^3_7(u,v)E_d(u)B_d(v)+ b^3_8(u,v,d)E_d(u)E_{\bar{d}}(v).
\end{eqnarray}
\begin{eqnarray}
D_a(u)B_b(v)&=& R^D_1(u,v)^{ac}_{db}B_c(v)D_d(u)+
c^1_1(u,v,\bar{a})B_{\bar{a}}(v)C_b(u)
+\delta_{ab}c^1_2(u,v)F(v)G(u)\nonumber \\
& & + c^1_3(u,v,\bar{a})B_{\bar{a}}(u)C_b(v)+c^1_4(u,v)E_a(u)C_b(v)+\delta_{ab}c^1_5(u,v)A(v)A(u)\nonumber \\
& & +c^1_6(u,v)A(v)\tilde{A}_{ab}(u)+\delta_{ab}c^1_7(u,v)A(u)A(v)+c^1_8(u,v)A(u)\tilde{A}_{ab}(v)\nonumber \\
& & +c^1_{9}(u,v)\tilde{A}_{ab}(u)A(v)+c^1_{10}(u,v)\tilde{A}_{ad}(u)\tilde{A}_{db}(v),\label{c1b1}\\[4mm]
C_a(u)B_b(v)&=& R^C_1(u,v)^{dc}_{ba}B_d(v)C_c(u)+
R^C_2(u,v)^{\bar{a}c}_{db}B_c(v)D_d(u)+\delta_{a\bar{b}} c^2_1(u,v,\bar{a})F(v)G(u)\nonumber \\
& &
+c^2_2(u,v)B_a(u)C_b(v)+c^2_3(u,v,\bar{a})E_{\bar{a}}(u)C_b(v)+\delta_{a\bar{b}}
c^2_4(u,v,\bar{a})A(v)A(u) \nonumber \\
 & &+
R^C_3(u,v)^{dc}_{ba}A(v)\tilde{A}_{\bar{d}c}(u)
+\delta_{a\bar{b}} c^2_5(u,v,\bar{a})A(v)\tilde{A}_2(u)\nonumber \\
& & +\delta_{a\bar{b}} c^2_6(u,v,\bar{a})A(u)A(v)+
c^2_{7}(u,v,\bar{a})A(u)\tilde{A}_{\bar{a}b}(v)\nonumber \\
& & + R^C_4(u,v)^{dc}_{ba}\tilde{A}_{\bar{d}c}(u)A(v)+
R^C_5(u,v)^{dc}_{ea}\tilde{A}_{\bar{d}c}(u)\tilde{A}_{eb}(v)\nonumber \\
& &+\delta_{a\bar{b}}c^2_{8}(u,v,\bar{a})\tilde{A}_2(u)A(v)+
c^2_{9}(u,v,\bar{a})\tilde{A}_2(u)\tilde{A}_{\bar{a}b}(v),\label{c2b1}
\end{eqnarray}
\begin{eqnarray}
B_a(u)E_b(v)&=& R^{be}_1(u,v)^{ca}_{bd}E_c(v)B_d(u)+
R^{be}_2(u,v)^{dc}_{\bar{b}a}B_d(v)B_c(u)
+\delta_{ab}e^1_1(u,v,a)F(v)A(u)\nonumber \\
& &+
R^{be}_3(u,v)^{dc}_{\bar{b}a}F(v)\tilde{A}_{\bar{d}c}(u)+\delta_{ab}e^1_2(u,v,a)F(u)A(v)
+R^{be}_4(u,v)^{ca}_{bd}F(u)\tilde{A}_{cd}(v)\nonumber \\
& & +\delta_{ab}e^1_3(u,v,a)F(u)\tilde{A}_2(v),\label{b1b2}
\end{eqnarray}
where  the coefficients $R^{\alpha}_i(u,v)^{ab}_{cd}$ are zeroes
except for $a=b=c=d, a=c\ne b=d, a=d\ne b=c$ and $a+b=c+d=q-1$,
$(\alpha=A,D,etc., i=1,2,etc.)$,
\begin{eqnarray}
&&g_1(u,v,a)
=-\frac{d_n(u_-,1,\bar{a}+1)b_n(2v)}{e_n(u_-)a_n(2v)},\hspace{4mm}
g_2(u,v,a) =\frac{d_n(u_+,1,\bar{a}+1)}{b_n(u_+)}\, \qquad
a+\bar{a}=q-1\nonumber .
\end{eqnarray}
\noindent The $\hat{r}(u),\tilde{r}(u)$ and $\bar{r}(u)$  are
given by
\begin{eqnarray}
&&\hat{r}(u)=\frac{1}{e_n(u)}\frac{b_n(u)}{a_n(u)}R^{(n-1)}(u),\hspace{2mm}\tilde{r}(u)
=\frac{1}{a_n(u)}R^{(n-1)}(u-4\eta),\hspace{2mm}\bar{r}(u)=\frac{1}{e_n(u)}R^{(n-1)}(u),\nonumber
\end{eqnarray}
respectively. The other coefficients are omitted here for their
long and tedious expressions.

\subsection{The m-particle state}

 \noindent Inferred from the commutation relation Eq.(\ref{b1b1}),
 we can construct the general m-particle state for these models as follow. Let
\begin{eqnarray}
& & \hspace{-6mm}\Phi_m^{b_1\cdots b_m}(v_1,\cdots,v_m)= B_{b_1}(v_1)\Phi_{m-1}^{b_2\cdots b_m}(v_2,\cdots,v_m)\nonumber \\
& & +F(v_1)\sum_{i=2}^m \Phi_{m-2}^{d_3\cdots
d_m}(v_2,\cdots,\check{v}_i,\cdots,v_m) S^{d_2\cdots
d_m}_{b_2\cdots b_m}(v_i;\{\check{v}_1,\check{v}_i\})\nonumber \\
& & \hspace{3mm}\times \Lambda_1^{m-2}(v_i;
\{\check{v}_1,\check{v}_i\})g_{1}(v_1,v_i,b_1)A(v_i)
\delta_{\bar{b}_1d_2}\nonumber \\
& & +F(v_1)\sum_{i=2}^m \Phi_{m-2}^{d_3\cdots
d_m}(v_2,\cdots,\check{v}_i,\cdots,v_m)
[\tilde{T}^{m-2}(v_i;\{\check{v}_1,\check{v}_i\})^{d_3\cdots
d_m}_{c_3\cdots c_m}]_{\bar{b}_1c_2}\nonumber \\
& & \hspace{3mm}\times S^{c_2\cdots c_m}_{b_2\cdots
b_m}(v_i;\{\check{v}_1,\check{v}_i\})g_2(v_1,v_i,b_1),
\label{nps2}
\end{eqnarray}
\noindent where
\begin{eqnarray}
S^{d_1\cdots d_m}_{b_1\cdots b_m}(v_i;\{\check{v}_i\})&=&
\hat{r}^{d_1d_2}_{c_2b_1}(v_1-v_i)\hat{r}^{c_2d_3}_{c_3b_2}(v_2-v_i)\cdots
\hat{r}^{c_{i-1}d_i}_{b_ib_{i-1}}(v_{i-1}-v_i)\prod^m_{j=i+1}\delta_{d_jb_j}
\nonumber \\[5mm] [\tilde{T}^{m}(u;\{{v}_m\})^{d_1\cdots
d_m}_{c_1\cdots c_m}]_{ab}&= &
\tilde{r}^{ad_1}_{h_1g_1}(u+v_1)\tilde{r}^{h_1d_2}_{h_2g_2}(u+v_2)\cdots
\tilde{r}^{h_{m-1}d_m}_{h_mg_m}(u+v_m)\tilde{A}_{h_mf_m}(u)\nonumber\\
& &
\bar{r}^{g_mf_m}_{c_mf_{m-1}}(u-v_m)\bar{r}^{g_{m-1}f_{m-1}}_{c_{m-1}f_{m-2}}(u-v_{m-1})
\cdots \bar{r}^{g_1f_1}_{c_1b}(u-v_1)
\end{eqnarray}
with  $ S^{d_1\cdots d_m}_{b_1\cdots
b_m}(v_1;v_2,\cdots,v_m)=\prod_{i=1}^{m}\delta_{d_ib_i}$,
$[\tilde{T}^{0}(u)]_{ab}=\tilde{A}_{ab}(u)$,
$\Lambda_l^m(u;v_1,v_2,\cdots,v_m)= \prod_{i=1}^m a^l_1(u,v_i)$, $
(l=1,3)$, $\Phi_0=1, \Phi_1^{b_1}(v_1)=B_{b_1}(v_1)$. The
$\check{v}_i$ means missing of ${v}_i$ in the sequence.

Then the general m-particle state is defined by
\begin{eqnarray}
|\Upsilon_m(v_1,\cdots,v_m)\rangle=\Phi_m^{b_1\cdots
b_m}(v_1,\cdots,v_m)F^{b_1\cdots b_m}|0\rangle, \label{nps1}
\end{eqnarray}
which  satisfy the $n-1$ exchange conditions
\begin{eqnarray}
& & \hspace{-6mm}\Phi_m^{b_1\cdots b_ib_{i+1}\cdots
b_m}(v_1,\cdots,v_i,v_{i+1},\cdots,v_m)F^{b_1\cdots b_m}|0\rangle= \nonumber\\
& & \Phi_m^{b_1\cdots a_ia_{i+1}\cdots
b_m}(v_1,\cdots,v_{i+1},v_{i},\cdots,v_m)\hat{r}^{a_ia_{i+1}}_{b_{i+1}b_i}(v_i-v_{i+1})F^{b_1\cdots
b_m}|0\rangle\ . \label{npsp}
\end{eqnarray}
We can prove Eq.(\ref{npsp}) by mathematical induction method.


\subsection{The eigenvalue and Bethe equations}

We can apply  the operators $x$ ($x=A,\tilde{A}_{aa},\tilde{A}_2$)
on the eigenstate ansatz   and obtain (see Appendix B)

\begin{eqnarray}
& &
\hspace{-10mm}{x}(u)|\Upsilon_m(v_1,\cdots,v_m)\rangle=|\tilde{\Psi}_{x}(u,\{v_m\})\rangle\nonumber\\
 & &+\sum_{i=1}^m h_1^x(u,v_i,d)|\Psi_{m-1}^{(1)}(u,v_i;\{v_m\})_{dd}\rangle\nonumber\\
& & +\sum_{i=1}^m  h_2^x(u,v_i,d)|\tilde{\Psi}_{m-1}^{(2)}(u,v_i;\{v_m\})_{dd}\rangle\nonumber\\
& &+\sum_{i=1}^m h_3^x(u,v_i,\bar{\alpha}_x)|\Psi_{m-1}^{(3)}(u,v_i;\{v_m\})_{\alpha_x\alpha_x}\rangle\nonumber\\
& &+\sum_{i=1}^m h_4^x(u,v_i,\bar{\alpha}_x)|\tilde{\Psi}_{m-1}^{(4)}(u,v_i;\{v_m\})_{\alpha_x\alpha_x}\rangle\nonumber\\
& & +\sum^{m-1}_{i=1}\sum_{j=i+1}^m
\tilde{H}^{x}_{1,d_1}(u,v_i,v_j)|\tilde{\Psi}_{m-2}^{(5)}(u,v_i,v_j;\{v_m\})_{d_1}\rangle\nonumber\\
& & +\sum^{m-1}_{i=1}\sum_{j=i+1}^m
\tilde{H}^{x}_{2,d_1}(u,v_i,v_j)|\tilde{\Psi}_{m-2}^{(6)}(u,v_i,v_j;\{v_m\})_{d_1}\rangle\nonumber\\
& & +\sum^{n-1}_{i=1}\sum_{j=i+1}^m
\tilde{H}^{x}_{3,d_1}(u,v_i,v_j)|\tilde{\Psi}_{m-2}^{(7)}(u,v_i,v_j;\{v_m\})_{d_1}\rangle\nonumber\\
& & +\sum^{m-1}_{i=1}\sum_{j=i+1}^m
\tilde{H}^{x}_{4,d_1}(u,v_i,v_j)|\tilde{\Psi}_{m-2}^{(8)}(u,v_i,v_j;\{v_m\})_{d_1}\rangle,
\label{nd1pns}
\end{eqnarray}
where  the expression of $|\tilde{\Psi}\rangle$'s and coefficients
$\tilde{H}^{x}_{j,d_1}$ ($j=1,2,3,4$) are  given in Appendix B.
 Using  Eq.(\ref{tru1}), we then get
\begin{eqnarray}
& & \hspace{-15mm}t(u)|\Upsilon_m(v_1,\cdots,v_m)\rangle
=w_1(u)\omega_1(u)\Lambda_1^m(u;v_1,\cdots,v_m)|\Upsilon_m(v_1,\cdots,v_m)\rangle\nonumber\\
& &+ {w}(u)\omega(u)\Lambda_2^m(u;v_1,\cdots,v_m)\Phi_m^{d_1\cdots
d_m}(v_1,\cdots,v_m)\tau_1(\tilde{u};\{\tilde{v}_m\})^{d_1\cdots
d_m}_{b_1\cdots b_m} F^{b_1\cdots
b_m}|0\rangle\nonumber\\
&
&+w_{q}(u)\omega_{q}(u)\Lambda_3^m(u;v_1,\cdots,v_m)|\Upsilon_m(v_1,\cdots,v_m)\rangle+
u.t. , \label{tupn}
\end{eqnarray}
where $u.t.$ denotes the unwanted terms,
\begin{eqnarray}
& & \hspace{-10mm}\tau_1(\tilde{u};\{\tilde{v}_m\})^{d_1\cdots
d_m}_{c_1\cdots
c_m}=\nonumber\\
& &
k^+({u})_a{L}(\tilde{u},\tilde{v}_1)^{ad_1}_{h_1g_1}{L}(\tilde{u},\tilde{v}_2)^{h_1d_2}_{h_2g_2}\cdots
{L}(\tilde{u},\tilde{v}_m)^{h_{m-1}d_m}_{h_mg_m}k^-({u})_{h_m}\nonumber\\
& & \times {L^{-1}}(-\tilde{u},\tilde{v}_m)^{h_mg_m}_{f_{m-1}c_m}
{L^{-1}}(-\tilde{u},\tilde{v}_{m-1})^{f_{m-1}g_{m-1}}_{f_{m-2}c_{m-1}}
\cdots
{L^{-1}}(-\tilde{u},\tilde{v}_1)^{f_1g_1}_{ac_1}.\label{taud}
\end{eqnarray}
with  $\tilde{v_i}=v_i-2\eta$, and
\begin{eqnarray}
& &
L(\tilde{u},\tilde{v})^{ab}_{cd}=R^{(n-1)}(\tilde{u}+\tilde{v})^{ab}_{cd},\nonumber\\
& &
L^{-1}(-\tilde{u},\tilde{v})^{ab}_{cd}=\frac{R^{(n-1)}(\tilde{u}-\tilde{v})^{ba}_{dc}}{\rho_{n-1}(\tilde{u}-\tilde{v})}.
\end{eqnarray}
Thus, we get the conclusion that$
|\Upsilon_m(v_1,\cdots,v_m)\rangle$ is the eigenstate of $t(u)$,
i.e.
\begin{eqnarray} & &
\hspace{-15mm}t(u)|\Upsilon_m(v_1,\cdots,v_m)\rangle
=\{w_1(u)\omega_1(u)\Lambda_1^m(u;v_1,\cdots,v_m)\nonumber\\
& &+ {w}(u)\omega(u)\Lambda_2^m(u;v_1,\cdots,v_m)
\Gamma_{1}(\tilde{u};\{\tilde{v}_{m}\};\{{v}_{m_{1}}^{(1)}\})\nonumber\\
&
&+w_{q}(u)\omega_{q}(u)\Lambda_3^m(u;v_1,\cdots,v_m)\}|\Upsilon_m(v_1,\cdots,v_m)\rangle \nonumber\\
& & =\Gamma(u;\{{v}_m\})|\Upsilon_m(v_1,\cdots,v_m)\rangle,
\label{tupn1}
\end{eqnarray} if the parameters satisfy
\begin{equation}
\tau_1(\tilde{u};\{\tilde{v}_m\}) F^{b_1\cdots
b_m}=\Gamma_{1}(\tilde{u};\{\tilde{v}_{m}\};\{{v}_{m_{1}}^{(1)}\})F^{b_1\cdots
b_m}\ , \label{tau1}
\end{equation}
\begin{equation}
\Gamma_{1}(\tilde{v}_i;\{\tilde{v}_{m}\};\{{v}_{m_{1}}^{(1)}\})=-\rho^{-\frac{1}{2}
}_{n-1}(0)\frac{\omega_1(v_i)\Lambda_1^{m-1}(v_i;
\{\check{v}_i\})} {\omega(v_i)\Lambda_2^{m-1}(v_i;
\{\check{v}_i\})}\beta_1(v_i).\hspace{2mm}(i=1,\cdots,m)\
\label{BA1}
\end{equation}
All unwanted terms  cancel out  by the following three kinds of
identities
\begin{eqnarray}
&&\beta_1(v_i)=T^{(d_1)}(v_i)\frac{w_1(u)a^1_2(u,v_i)+\sum_{d=1}^{q-2}w_{{d}+1}(u)R^A_1(u,v_i)^{dd_1}_{d_1d}
+w_{q}(u)a^3_2(u,v_i)}
{w_1(u)a^1_3(u,v_i)+\sum_{d=1}^{q-2}w_{{d}+1}(u)R^A_2(u,v_i)^{dd_1}_{d_1d}+w_{q}(u)a^3_3(u,v_i)},\label{bat1}\\[4mm]
&&\beta_1(v_i)=T^{(d_1)}(v_i)\frac{w_{\bar{d}_1+1}(u)R^A_3(u,v_i,d_1)+w_{q}(u)a^3_4(u,v_i,d_1)}
{w_{\bar{d}_1+1}(u)R^A_4(u,v_i,d_1)+w_{q}(u)a^3_5(u,v_i,d_1)},\label{bat2}\\[4mm]
&&\sum_{l=1}^{q}
w_l(u)\tilde{H}^{x_l}_{1,d_1}(u,v_i,v_j)-[\sum_{l=1}^{q}
w_l(u)\tilde{H}^{x_l}_{2,d_1}(u,v_i,v_j)]\beta_1(v_i)\nonumber\\
&&-[\sum_{l=1}^{q}
w_l(u)\tilde{H}^{x_l}_{3,d_1}(u,v_i,v_j)]\beta_1(v_j)+
[\sum_{l=1}^{q}
w_l(u)\tilde{H}^{x_l}_{4,d_1}(u,v_i,v_j)]\beta_1(v_i)\beta_1(v_j)=0\label{bat3}
\end{eqnarray}
with $x_1=A, x_{l+1}=\tilde{A}_{ll}, x_{q}=\tilde{A}_2$,
$w_{d+1}(u)=w(u)k^+_{d}(u)$, $d_1=1,2,\cdots, q-2$ do not sum in
Eqs.(\ref{bat1},\ref{bat2},\ref{bat3}).

From Eqs.(\ref{taud}), (\ref{tupn1}) and (\ref{tau1}) , we can see
 that the diagonalization of $\tau(u)$ is reduced to finding
 the eigenvalue of $\tau_1(\tilde{u};\{\tilde{v}_m\})$  which is just the transfer matrix of the same open model with
 its $R$ matrix given by  $R^{(n-1)}(u)$.
 Repeating the
procedure $j$ times, we can obtain the eigenvalue
$\Gamma_j({u}^{(j)};\{\tilde{v}_{m_{j-1}}^{(j-1)}\};\{{v}_{m_{j}}^{(j)}\})$
with its transfer matrix constructed by  $R^{(n-j)}(u)$,
\begin{eqnarray}
& & \hspace{-5mm}
\Gamma_j({u}^{(j)};\{\tilde{v}_{m_{j-1}}^{(j-1)}\};\{{v}_{m_{j}}^{(j)}\})\nonumber\\
&&=
w_1^{(j)}(u^{(j)})\omega_1^{(j)}(u^{(j)};\{\tilde{v}_{m_{j-1}}^{(j-1)}\}
)\Lambda_1^{m_j}(u^{(j)};\{{v}_{m_{j}}^{(j)}\})\nonumber\\
& &+
{w}^{(j)}(u^{(j)})\omega^{(j)}(u^{(j)};\{\tilde{v}_{m_{j-1}}^{(j-1)}\})
\Lambda_2^{m_j}(u^{(j)};\{{v}_{m_{j}}^{(j)}\})
\Gamma_{j+1}(u^{(j+1)};\{\tilde{v}_{m_{j}}^{(j)}\};\{{v}_{m_{j+1}}^{(j+1)}\})\nonumber\\
&
&+w_{q-2j}^{(j)}(u^{(j)})\omega_{q-2j}^{(j)}(u^{(j)};\{\tilde{v}_{m_{j-1}}^{(j-1)}\})
\Lambda_3^{m_j}(u^{(j)};\{{v}_{m_{j}}^{(j)}\}) \label{nesttu}
\end{eqnarray}
with ${u}^{(j)}={u}-{2j}\eta$,
 $\tilde{v}_{k}^{(j)}={v}_{k}^{(j)}-2\eta$,
 $\{{v}_{m_{j}}^{(j)}\}=\{v_1^{(j)}, \cdots,v_{m_j}^{(j)}\}$, $\{{v}_{m_{0}}^{(0)}\}= \{{v}_{m}\}$,
 $\{{v}_{m_{-1}}^{(-1)}\}=\{\tilde{v}_{m_{-1}}^{(-1)}\}= \{0\}$, $m_{-1}=N,
 m_0=m$,

\begin{eqnarray}
&&\omega_1^{(j)}(u^{(j)};\{\tilde{v}_{m_{j-1}}^{(j-1)}\})=
\bar{\omega}_1^{(j)}(u^{(j)})\xi_1^{(j)}(u^{(j)};\{\tilde{v}_{m_{j-1}}^{(j-1)}\}),\nonumber\\
&&\omega^{(j)}(u^{(j)};\{\tilde{v}_{m_{j-1}}^{(j-1)}\})=
\bar{\omega}^{(j)}(u^{(j)})\xi_2^{(j)}(u^{(j)};\{\tilde{v}_{m_{j-1}}^{(j-1)}\}),\nonumber\\
&&\omega_{q-2j}^{(j)}(u^{(j)};\{\tilde{v}_{m_{j-1}}^{(j-1)}\})=
\bar{\omega}_{q-2j}^{(j)}(u^{(j)})\xi_3^{(j)}(u^{(j)};\{\tilde{v}_{m_{j-1}}^{(j-1)}\})
\end{eqnarray}
\begin{eqnarray}
&&\xi_1^{(j)}(u^{(j)};\{\tilde{v}_{m_{j-1}}^{(j-1)}\})=
\prod_{i=1}^{m_{j-1}}\frac{a_{n-j}(u^{(j)}+\tilde{v}^{(j-1)}_i)a_{n-j}(u^{(j)}-\tilde{v}^{(j-1)}_i)}
{a_{n-j}(u^{(j)}-\tilde{v}^{(j-1)}_i)a_{n-j}(\tilde{v}^{(j-1)}_i-u^{(j)})},\nonumber\\
&&\xi_2^{(j)}(u^{(j)};\{\tilde{v}_{m_{j-1}}^{(j-1)}\})=
\prod_{i=1}^{m_{j-1}}\frac{b_{n-j}(u^{(j)}+\tilde{v}^{(j-1)}_i)b_{n-j}(u^{(j)}-\tilde{v}^{(j-1)}_i)}
{a_{n-j}(u^{(j)}-\tilde{v}^{(j-1)}_i)a_{n-j}(\tilde{v}^{(j-1)}_i-u^{(j)})},\nonumber\\
&&\xi_3^{(j)}(u^{(j)};\{\tilde{v}_{m_{j-1}}^{(j-1)}\})=
\prod_{i=1}^{m_{j-1}}\frac{e_{n-j}(u^{(j)}+\tilde{v}^{(j-1)}_i)e_{n-j}(u^{(j)}-\tilde{v}^{(j-1)}_i)}
{a_{n-j}(u^{(j)}-\tilde{v}^{(j-1)}_i)a_{n-j}(\tilde{v}^{(j-1)}_i-u^{(j)})}.
\end{eqnarray}

\begin{eqnarray}
\Lambda_1^{m_j}(u^{(j)};\{{v}_{m_{j}}^{(j)}\})&=&\prod_{i=1}^{m_j}
\frac{a_{n-j}(-\tilde{u}^{(j)}+\tilde{v}_i^{(j)})a_{n-j}(-\tilde{u}^{(j)}-\tilde{v}_i^{(j)})}
{b_{n-j}(-\tilde{u}^{(j)}+\tilde{v}_i^{(j)})b_{n-j}(-\tilde{u}^{(j)}-\tilde{v}_i^{(j)})}\nonumber\\
\Lambda_2^{m_j}(u^{(j)};\{{v}_{m_{j}}^{(j)}\})&=&\prod_{i=1}^{m_j}
\frac{a_{n-j-1}(-\tilde{u}^{(j)}+\tilde{v}_i^{(j)})a_{n-j-1}(\tilde{u}^{(j)}-\tilde{v}_i^{(j)})}
{e_{n-j}(\tilde{u}^{(j)}-\tilde{v}_i^{(j)})e_{n-j}(\tilde{u}^{(j)}+\tilde{v}_i^{(j)})}\nonumber\\
\Lambda_3^{m_j}(u^{(j)};\{{v}_{m_{j}}^{(j)}\})&=&\prod_{i=1}^{m_j}
\frac{b_{n-j}(\tilde{u}^{(j)}-\tilde{v}_i^{(j)})b_{n-j}(\tilde{u}^{(j)}+\tilde{v}_i^{(j)})}
{e_{n-j}(\tilde{u}^{(j)}-\tilde{v}_i^{(j)})e_{n-j}(\tilde{u}^{(j)}+\tilde{v}_i^{(j)})}.\nonumber
\end{eqnarray}
The  coefficients $w$'s, $\bar{\omega}$'s vary with the
corresponding transfer matrix.
 The Bethe anzatz equations are
\begin{eqnarray}
&&\hspace{-8mm}\Gamma_{j+1}(\tilde{v}_{i}^{(j)};\{\tilde{v}_{m_{j}}^{(j)}\};\{{v}_{m_{j+1}}^{(j+1)}\})\nonumber\\
&&=-\rho^{-\frac{1}{2}
}_{n-j-1}(0)\frac{\omega_1^{(j)}(v^{(j)}_i;\{\tilde{v}_{m_{j-1}}^{(j-1)}\}
)\Lambda_1^{m_j-1}(v^{(j)}_i;\{\check{v}_{i}^{(j)}\})}
{\omega^{(j)}(v^{(j)}_i;\{\tilde{v}_{m_{j-1}}^{(j-1)}\}
)\Lambda_2^{m_j-1}(v^{(j)}_i;\{\check{v}_{i}^{(j)}\})}\beta_{j+1}(v_i^{(j)}).
\hspace{2mm}(i=1,\cdots,m_j)\label{nestBA}
\end{eqnarray}
When $j=n-1$, our problem become to obtain  the eigenvalues of
 the nineteen-vertex models $A^{(2)}_{2}$,
 $B^{(1)}_1$ for
 $A^{(2)}_{2n}$, $B^{(1)}_n$  and six-vertex models $A^{(2)}_{1},C^{(1)}_1$ for $A^{(2)}_{2n-1},C^{(1)}_n$, respectively.
 For $D^{(1)}_n$ when $j=n-2$, we will arrive at $D^{(1)}_2$
 model whose eigenvalues  can be given in terms of the product of the eigenvalues of
 two $A^{(1)}_1$  vertex models\cite{10},\cite{12},\cite{lima}.  The six-vertex and  nineteen-vertex models have been well
 solved\cite{ik},\cite{osp12}.
 Thus, we can get  the whole eigenvalues and the
Bethe ansatz equations of transfer matrix   for the
$A^{(2)}_{2n},A^{(2)}_{2n-1},B^{(1)}_n,C^{(1)}_n,D^{(1)}_{n}$
vertex model with open boundary conditions.

From Eq.(\ref{tupn1}),  we can obtain the energy of the
Hamiltonian Eq.(\ref{hh})
\begin{equation}
E=\frac{1}{2trK_{+}(0)}[\frac{\Gamma'(0;\{{v}_m\})}{c_{\zeta}}-trK'_{+}(0)].\label{energy}
\end{equation}
The more explicit expression for Eq.(\ref{energy}) is
\begin{eqnarray}
E&=&\frac{1}{2trK_{+}(0)}[w'_1(0)+w_1(0)\frac{\bar{\omega}'_1(0)}{c_{\zeta}}
+2Nw_1(0)\frac{a'_n(0)}{a_n(0)}\nonumber\\
&&-w_1(0)\sum^N_{i=1}\frac{2\sinh(2\eta)}{\cosh(v_i-2\eta)-\cosh(2\eta)}-trK'_{+}(0)].\label{energy1}
\end{eqnarray}

\subsection{ The quantum-algebra-invariant cases}

Here we will consider  the trivial K matrices for
$A^{(2)}_{2n},A^{(2)}_{2n-1},B^{(1)}_n,C^{(1)}_n,D^{(1)}_{n}$
models (quantum-algebra-invariant case). The K matrices take the
forms
\begin{equation}
K_{-}=1,\hspace{5mm}K_{+}=M(n).
\end{equation}
We begin with $A^{(2)}_{2n},B^{(1)}_n$ models. After a not very
long derivation, we can rewrite Eq.(\ref{tupn1}) in detail, which
is
\begin{eqnarray}
& & \hspace{-5mm} \Gamma({u})=\Gamma_0({u})=
{w}_1^{(0)}(u^{(0)})\bar{\omega}_1^{(0)}(u^{(0)} )
\xi_1^{(0)}(u^{(0)};\{\tilde{v}_{m_{-1}}^{(-1)}\}){\cal{A}}^{(m_0)}(u)\nonumber\\
&&+{w}_{q}^{(0)}(u^{(0)})\bar{\omega}_{q}^{(0)}(u^{(0)} )
\xi_3^{(0)}(u^{(0)};\{\tilde{v}_{m_{-1}}^{(-1)}\}){\cal{C}}^{(m_0)}(u)\nonumber\\
&&
+\sum_{j=0}^{n-2}\mu_j(u^{(j)})\nu_j(u^{(j)}){w}_1^{(j+1)}(u^{(j+1)})\bar{\omega}_1^{(j+1)}(u^{(j+1)}
)\xi_2^{(0)}(u^{(0)};\{\tilde{v}_{m_{-1}}^{(-1)}\}){\cal{B}}^{(m_j,m_{j+1})}(u)\nonumber\\
&&
+\sum_{j=0}^{n-2}\mu_j(u^{(j)})\nu_j(u^{(j)}){w}_{q-2(j+1)}^{(j+1)}(u^{(j+1)})\bar{\omega}_{q-2(j+1)}^{(j+1)}(u^{(j+1)}
)\xi_2^{(0)}(u^{(0)};\{\tilde{v}_{m_{-1}}^{(-1)}\})\bar{\cal{B}}^{(m_j,m_{j+1})}(u)\nonumber\\
&&+\mu_{n-1}(u^{(n-1)})\nu_{n-1}(u^{(n-1)})\xi_2^{(0)}(u^{(0)};\{\tilde{v}_{m_{-1}}^{(-1)}\}){\cal{{B}}}^{(m_{n-1})}(u),\label{nesttu11}
\end{eqnarray}
where the identity
$\Lambda_2^{m_j}(u^{(j)};\{{v}_{m_{j}}^{(j)}\})\xi_2^{(j+1)}(u^{(j+1)};\{\tilde{v}_{m_{j}}^{(j)}\})=1
$ is used,
\begin{equation}
\mu_j(u^{(j)})=\prod_{i=0}^{j}\bar{\omega}^{(i)}(u^{(i)}),
\hspace{4mm}\nu_j(u^{(j)})=\prod_{i=0}^{j}{w}^{(i)}(u^{(i)}),
\end{equation}
\begin{eqnarray}
&&{\cal{A}}^{(m_0)}(u)=\prod_{k=1}^{m_0}
\frac{a_{n}(-\tilde{u}^{(0)}+\tilde{v}_k^{(0)})a_{n}(-\tilde{u}^{(0)}-\tilde{v}_k^{(0)})}
{b_{n}(-\tilde{u}^{(0)}+\tilde{v}_k^{(0)})b_{n}(-\tilde{u}^{(0)}-\tilde{v}_k^{(0)})},\label{A}\\[4mm]
&&{\cal{C}}^{(m_0)}(u)=\prod_{k=1}^{m_0}
\frac{b_{n}(\tilde{u}^{(0)}-\tilde{v}_k^{(0)})b_{n}(\tilde{u}^{(0)}+\tilde{v}_k^{(0)})}
{e_{n}(\tilde{u}^{(0)}-\tilde{v}_k^{(0)})e_{n}(\tilde{u}^{(0)}+\tilde{v}_k^{(0)})},\label{C}\\[4mm]
&&{\cal{B}}^{(m_j,m_{j+1})}(u)=\prod_{k=1}^{m_j}
\frac{a_{n-j-1}(\tilde{u}^{(j)}+\tilde{v}_k^{(j)})a_{n-j-1}(\tilde{u}^{(j)}-\tilde{v}_k^{(j)})}
{e_{n-j}(\tilde{u}^{(j)}+\tilde{v}_k^{(j)})e_{n-j}(\tilde{u}^{(j)}-\tilde{v}_k^{(j)})}\nonumber\\
&&\hspace{30mm}\times\prod_{l=1}^{m_{j+1}}
\frac{a_{n-j-1}(-\tilde{u}^{(j+1)}+\tilde{v}_l^{(j+1)})a_{n-j-1}(-\tilde{u}^{(j+1)}-\tilde{v}_l^{(j+1)})}
{b_{n-j-1}(-\tilde{u}^{(j+1)}+\tilde{v}_l^{(j+1)})b_{n-j-1}(-\tilde{u}^{(j+1)}-\tilde{v}_l^{(j+1)})},\label{Bmm1}\\[4mm]
&&\bar{\cal{{B}}}^{(m_j,m_{j+1})}(u)=\prod_{k=1}^{m_j}
\frac{e_{n-j-1}(\tilde{u}^{(j)}+\tilde{v}_k^{(j)})e_{n-j-1}(\tilde{u}^{(j)}-\tilde{v}_k^{(j)})}
{e_{n-j}(\tilde{u}^{(j)}+\tilde{v}_k^{(j)})e_{n-j}(\tilde{u}^{(j)}-\tilde{v}_k^{(j)})}\nonumber\\
&&\hspace{30mm}\times\prod_{l=1}^{m_{j+1}}
\frac{b_{n-j-1}(\tilde{u}^{(j+1)}-\tilde{v}_l^{(j+1)})b_{n-j-1}(\tilde{u}^{(j+1)}+\tilde{v}_l^{(j+1)})}
{e_{n-j-1}(\tilde{u}^{(j+1)}-\tilde{v}_l^{(j+1)})e_{n-j-1}(\tilde{u}^{(j+1)}+\tilde{v}_l^{(j+1)})},\label{Bmm2}\\
&& \hspace{10cm}(j=0,1,2,\cdots,n-2)\nonumber
\end{eqnarray}
\begin{eqnarray}
&&{\cal{{B}}}^{(m_{n-1})}(u)=\prod_{k=1}^{m_{n-1}}
\frac{f_{0}(\tilde{u}^{(n-1)}+\tilde{v}_k^{(n-1)})f_{0}(\tilde{u}^{(n-1)}-\tilde{v}_k^{(n-1)})}
{e_{1}(\tilde{u}^{(n-1)}+\tilde{v}_k^{(n-1)})e_{1}(\tilde{u}^{(n-1)}-\tilde{v}_k^{(n-1)})},\nonumber
\end{eqnarray}
where $f_0(u)$  is just the matric elements  $f_n(u)$ when $n=0$.
 The more explicit expression of Bethe equations Eq.(\ref{nestBA})
are
\begin{eqnarray}
\displaystyle&&\prod_{k=1}^{m_{j-1}}\frac{a_{n-j}(\tilde{v}^{(j)}_i+\tilde{v}^{(j-1)}_k)
a_{n-j}(\tilde{v}^{(j)}_i-\tilde{v}^{(j-1)}_k)}
{b_{n-j}(\tilde{v}^{(j)}_i+\tilde{v}^{(j-1)}_k)b_{n-j}(\tilde{v}^{(j)}_i-\tilde{v}^{(j-1)}_k)}\nonumber\\
&&\times\prod_{l=1}^{m_{j+1}}\frac{b_{n-j-1}(-\tilde{v}^{(j)}_i+\tilde{v}^{(j+1)}_l)
b_{n-j-1}(-\tilde{v}^{(j)}_i-\tilde{v}^{(j+1)}_l)}
{a_{n-j-1}(-\tilde{v}^{(j)}_i+\tilde{v}^{(j+1)}_l)a_{n-j-1}(-\tilde{v}^{(j)}_i-\tilde{v}^{(j+1)}_l)}\nonumber\\
&&\times\prod_{s=1\ne i
}^{m_{j}}\frac{a_{n-j}(-\tilde{v}^{(j)}_i+\tilde{v}^{(j)}_s)
a_{n-j}(-\tilde{v}^{(j)}_i-\tilde{v}^{(j)}_s)}
{b_{n-j}(-\tilde{v}^{(j)}_i+\tilde{v}^{(j)}_s)b_{n-j}(-\tilde{v}^{(j)}_i-\tilde{v}^{(j)}_s)}
\frac{e_{n-j}(\tilde{v}^{(j)}_i+\tilde{v}^{(j)}_s)
e_{n-j}(\tilde{v}^{(j)}_i-\tilde{v}^{(j)}_s)}
{a_{n-j-1}(\tilde{v}^{(j)}_i+\tilde{v}^{(j)}_s)a_{n-j-1}(\tilde{v}^{(j)}_i-\tilde{v}^{(j)}_s)}\nonumber\\
&&=-\frac{{w}_1^{(j+1)}(\tilde{v}_i^{(j)})a_{n-j-1}(2\tilde{v}_i^{(j)}))
}{\beta_{j+1}(v_i^{(j)})}\frac{\bar{\omega}^{(j)}(v_i^{(j)})\bar{\omega}_1^{(j+1)}(\tilde{v}_i^{(j)})}
{\bar{\omega}_1^{(j)}({v}_i^{(j)})}, \hspace{2mm}(i=1,\cdots,m_j;
j\ne n-1)\label{nestBAB1}
\end{eqnarray}

\begin{eqnarray}
\displaystyle&&\prod_{k=1}^{m_{n-2}}\frac{a_{1}(\tilde{v}^{(n-1)}_i+\tilde{v}^{(n-2)}_k)
a_{1}(\tilde{v}^{(n-1)}_i-\tilde{v}^{(n-2)}_k)}
{b_{1}(\tilde{v}^{(n-1)}_i+\tilde{v}^{(n-2)}_k)b_{1}(\tilde{v}^{(n-1)}_i-\tilde{v}^{(n-2)}_k)}\nonumber\\
&&\times\prod_{l=1\ne i
}^{m_{n-1}}\frac{a_{1}(-\tilde{v}^{(n-1)}_i+\tilde{v}^{(n-1)}_l)
a_{1}(-\tilde{v}^{(n-1)}_i-\tilde{v}^{(n-1)}_l)}
{b_{1}(-\tilde{v}^{(n-1)}_i+\tilde{v}^{(n-1)}_l)b_{1}(-\tilde{v}^{(n-1)}_i-\tilde{v}^{(n-1)}_l)}
\frac{e_{1}(\tilde{v}^{(n-1)}_i+\tilde{v}^{(n-1)}_l)
e_{1}(\tilde{v}^{(n-1)}_i-\tilde{v}^{(n-1)}_l)}
{f_{0}(\tilde{v}^{(n-1)}_i+\tilde{v}^{(n-1)}_l)f_{0}(\tilde{v}^{(n-1)}_i-\tilde{v}^{(n-1)}_l)}\nonumber\\
&&=-\frac{\bar{\omega}^{(n-1)}(v_i^{(n-1)})}
{\bar{\omega}_1^{(n-1)}({v}_i^{(n-1)})\beta_{n}(v_i^{(n-1)})},
\label{nestBAB2} \hspace{2mm}(i=1,\cdots,m_{n-1})
\end{eqnarray}

From Eqs.(\ref{Avac}, \ref{A1vac}, \ref{A3vac}, \ref{A4vac}) we
can obtain
\begin{eqnarray}
&& \bar{\omega}_{1}^{(j)}(u^{(j)})=1,
\hspace{4mm}\bar{\omega}^{(j)}(u^{(j)})=\frac{e^{2\eta}\sinh(u^{(j)})}{\sinh(u^{(j)}-2\eta)},\nonumber
\\
&&
\bar{\omega}_{2(n-j)+1}^{(j)}(u^{(j)})=\frac{e^{2(2(n-j)-1)\eta}\sinh(u^{(j)})\cosh(u^{(j)}-(2(n-j)+3)\eta)}
{\sinh(u^{(j)}-4(n-j)\eta)\cosh(u^{(j)}-(2(n-j)+1)\eta)}.
\end{eqnarray}
\begin{eqnarray}
&&
w_{1}^{(j)}(u^{(j)})=\frac{\sinh(u^{(j)}-2(2(n-j)+1)\eta)\cosh(u^{(j)}-(2(n-j)-1)\eta)}
{\sinh(u^{(j)}-2\eta)\cosh(u^{(j)}-(2(n-j)+1)\eta)},\nonumber\\
&&
w^{(j)}(u^{(j)})=\frac{e^{-2\eta}\sinh(u^{(j)}-2(2(n-j)+1)\eta)}{\sinh(u^{(j)}-4(n-j)\eta)},
\hspace{4mm}w_{2(n-j)+1}^{(j)}(u^{(j)})=e^{-2(2(n-j)-1)\eta},
\end{eqnarray}
\begin{equation}
\beta_{j+1}(v_i^{(j)})=\displaystyle\left\{\begin{array}{ll}
\displaystyle-\frac{2e^{2\eta}\sinh(v_i^{(j)})\sinh(v_i^{(j)}-4(n-j)\eta)
\cosh(v_i^{(j)}-(2n-2j-1)\eta)}{\sinh(v_i^{(j)}-2\eta)},&
j\le n-2\\
\displaystyle-\frac{e^{2\eta}\sinh(v_i^{(n-1)})}{\sinh(v_i^{(n-1)}-2\eta)}&
j=n-1
\end{array}\right.
\end{equation}
 and
\begin{eqnarray}
&& \bar{\omega}_{1}^{(j)}(u^{(j)})=1,
\hspace{4mm}\bar{\omega}^{(j)}(u^{(j)})=\frac{e^{2\eta}\sinh(u^{(j)})}{\sinh(u^{(j)}-2\eta)},\nonumber
\\
&&
\bar{\omega}_{2(n-j)+1}^{(j)}(u^{(j)})=\frac{e^{2(2(n-j)-1)\eta}\sinh(u^{(j)})\sinh(u^{(j)}-(2(n-j)-3)\eta)}
{\sinh(u^{(j)}-4(n-j-1)\eta)\sinh(u^{(j)}-(2(n-j)-1)\eta)},\label{bnow}
\end{eqnarray}
\begin{eqnarray}
&&
w_{1}^{(j)}(u^{(j)})=\frac{\sinh(u^{(j)}-(2(n-j)+1)\eta)\sinh(u^{(j)}-2(2(n-j)-1)\eta)}
{\sinh(u^{(j)}-2\eta)\sinh(u^{(j)}-(2(n-j)-1)\eta)},\nonumber\\
&&
w^{(j)}(u^{(j)})=\frac{e^{-2\eta}\sinh(u^{(j)}-2(2(n-j)-1)\eta)}{\sinh(u^{(j)}-4(n-j-1)\eta)},
\hspace{4mm}w_{2(n-j)+1}^{(j)}(u^{(j)})=e^{-2(2(n-j)-1)\eta},\label{bnww}
\end{eqnarray}
\begin{equation}
\beta_{j+1}(v_i^{(j)})=\displaystyle\left\{\begin{array}{ll}
\displaystyle-\frac{2e^{2\eta}\sinh(v_i^{(j)})\sinh(v_i^{(j)}-4(n-j-1)\eta)
\sinh(v_i^{(j)}-(2n-2j+1)\eta)}{\sinh(v_i^{(j)}-2\eta)},&
j\le n-2\\
\displaystyle-\frac{e^{2\eta}\sinh(v_i^{(n-1)})}{\sinh(v_i^{(n-1)}-2\eta)}&
j=n-1
\end{array}\right.\label{bnbeta}
\end{equation}
for  $A^{(2)}_{2n}$ and $B^{(1)}_{n}$, respectively.
The Eq.(\ref{tupn1}) for $A^{(2)}_{2n-1},C^{(1)}_n$ and
$D^{(1)}_n$  can be written as
\begin{eqnarray}
& & \hspace{-5mm} \Gamma({u})=\Gamma_0({u})=
{w}_1^{(0)}(u^{(0)})\bar{\omega}_1^{(0)}(u^{(0)} )
\xi_1^{(0)}(u^{(0)};\{\tilde{v}_{m_{-1}}^{(-1)}\}){\cal{A}}^{(m_0)}(u)\nonumber\\
&&+{w}_{q}^{(0)}(u^{(0)})\bar{\omega}_{q}^{(0)}(u^{(0)} )
\xi_3^{(0)}(u^{(0)};\{\tilde{v}_{m_{-1}}^{(-1)}\}){\cal{C}}^{(m_0)}(u)\nonumber\\
&&
+\sum_{j=0}^{n-2}\mu_j(u^{(j)})\nu_j(u^{(j)}){w}_1^{(j+1)}(u^{(j+1)})\bar{\omega}_1^{(j+1)}(u^{(j+1)}
)\xi_2^{(0)}(u^{(0)};\{\tilde{v}_{m_{-1}}^{(-1)}\}){\cal{B}}^{(m_j,m_{j+1})}(u)\nonumber\\
&&
+\sum_{j=0}^{n-2}\mu_j(u^{(j)})\nu_j(u^{(j)}){w}_{q-2(j+1)}^{(j+1)}(u^{(j+1)})\bar{\omega}_{q-2(j+1)}^{(j+1)}(u^{(j+1)}
)\xi_2^{(0)}(u^{(0)};\{\tilde{v}_{m_{-1}}^{(-1)}\})\bar{\cal{B}}^{(m_j,m_{j+1})}(u)\label{nesttu12}
\end{eqnarray}
and
\begin{eqnarray}
& & \hspace{-5mm} \Gamma({u})=\Gamma_0({u})=
{w}_1^{(0)}(u^{(0)})\bar{\omega}_1^{(0)}(u^{(0)} )
\xi_1^{(0)}(u^{(0)};\{\tilde{v}_{m_{-1}}^{(-1)}\}){\cal{A}}^{(m_0)}(u)\nonumber\\
&&+{w}_{q}^{(0)}(u^{(0)})\bar{\omega}_{q}^{(0)}(u^{(0)} )
\xi_3^{(0)}(u^{(0)};\{\tilde{v}_{m_{-1}}^{(-1)}\}){\cal{C}}^{(m_0)}(u)\nonumber\\
&&
+\sum_{j=0}^{n-3}\mu_j(u^{(j)})\nu_j(u^{(j)}){w}_1^{(j+1)}(u^{(j+1)})\bar{\omega}_1^{(j+1)}(u^{(j+1)}
)\xi_2^{(0)}(u^{(0)};\{\tilde{v}_{m_{-1}}^{(-1)}\}){\cal{B}}^{(m_j,m_{j+1})}(u)\nonumber\\
&&
+\sum_{j=0}^{n-3}\mu_j(u^{(j)})\nu_j(u^{(j)}){w}_{q-2(j+1)}^{(j+1)}(u^{(j+1)})\bar{\omega}_{q-2(j+1)}^{(j+1)}(u^{(j+1)}
)\xi_2^{(0)}(u^{(0)};\{\tilde{v}_{m_{-1}}^{(-1)}\})\bar{\cal{B}}^{(m_j,m_{j+1})}(u)\nonumber\\
&&
+\mu_{n-3}(u^{(n-3)})\nu_{n-3}(u^{(n-3)}){w}_1^{(n-1)}(u^{(n-2)})\bar{\omega}_1^{(n-1)}(u^{(n-2)}
)\xi_2^{(0)}(u^{(0)};\{\tilde{v}_{m_{-1}}^{(-1)}\}){\cal{B}}^{(m_{n-2},m_{n-1})}(u)\nonumber\\
&&+\mu_{n-3}(u^{(n-3)})\nu_{n-3}(u^{(n-3)}){w}_{2}^{(n-1)}(u^{(n-2)})\bar{\omega}_{2}^{(n-1)}(u^{(n-2)}
)\xi_2^{(0)}(u^{(0)};\{\tilde{v}_{m_{-1}}^{(-1)}\})\bar{\cal{B}}^{(m_{n-2},m_{n-1})}(u),\label{nesttu13}
\end{eqnarray}
respectively. Here Eqs.(\ref{Bmm1},\ref{Bmm2},\ref{nestBAB1}) hold
until $j=n-3$ for $A^{(2)}_{2n-1},C^{(1)}_n$ and $j=n-4$ for
$D^{(1)}_n$. The rest $\cal{B}$'s and $\bar{\cal{B}}$'s are

\begin{eqnarray}
&&{\cal{B}}^{(m_{n-2},m_{n-1})}(u)=\prod_{k=1}^{m_{n-2}}
\frac{a_1(\tilde{u}^{(n-2)}+\tilde{v}_k^{(n-2)})a_1(\tilde{u}^{(n-2)}-\tilde{v}_k^{(n-2)})}
{e_{2}(\tilde{u}^{(n-2)}+\tilde{v}_k^{(n-2)})e_{2}(\tilde{u}^{(n-2)}-\tilde{v}_k^{(n-2)})}\nonumber\\
&&\hspace{30mm}\times\prod_{l=1}^{m_{n-1}}
\frac{a_1(-\tilde{u}^{(n-1)}+\tilde{v}_l^{(n-1)})a_1(-\tilde{u}^{(n-1)}-\tilde{v}_l^{(n-1)})}
{e_1(-\tilde{u}^{(n-1)}+\tilde{v}_l^{(n-1)})e_1(-\tilde{u}^{(n-1)}-\tilde{v}_l^{(n-1)})},\nonumber\\[4mm]
&&\bar{\cal{{B}}}^{(m_{n-2},m_{n-1})}(u)=\prod_{k=1}^{m_{n-2}}
\frac{e_1(\tilde{u}^{(n-2)}+\tilde{v}_k^{(n-2)})e_1(\tilde{u}^{(n-2)}-\tilde{v}_k^{(n-2)})}
{e_{2}(\tilde{u}^{(n-2)}+\tilde{v}_k^{(n-2)})e_{2}(\tilde{u}^{(n-2)}-\tilde{v}_k^{(n-2)})}\nonumber\\
&&\hspace{30mm}\times\prod_{l=1}^{m_{n-1}}
\frac{a_1(\tilde{u}^{(n-1)}+\tilde{v}_l^{(n-1)})a_1(\tilde{u}^{(n-1)}-\tilde{v}_l^{(n-1)})}
{e_1(\tilde{u}^{(n-1)}+\tilde{v}_l^{(n-1)})e_1(\tilde{u}^{(n-1)}-\tilde{v}_l^{(n-1)})}.\nonumber
\end{eqnarray}
for  $A^{(2)}_{2n-1}$,
\begin{eqnarray}
&&{\cal{B}}^{(m_{n-2},m_{n-1})}(u)=\prod_{k=1}^{m_{n-2}}
\frac{a_1(\tilde{u}^{(n-2)}+\tilde{v}_k^{(n-2)})a_1(\tilde{u}^{(n-2)}-\tilde{v}_k^{(n-2)})}
{e_{2}(\tilde{u}^{(n-2)}+\tilde{v}_k^{(n-2)})e_{2}(\tilde{u}^{(n-2)}-\tilde{v}_k^{(n-2)})}\nonumber\\
&&\hspace{30mm}\times\prod_{l=1}^{m_{n-1}}
\frac{a(-\tilde{u}^{(n-1)}+\tilde{v}_l^{(n-1)})b(\tilde{u}^{(n-1)}+\tilde{v}_l^{(n-1)}+4\eta)}
{b(-\tilde{u}^{(n-1)}+\tilde{v}_l^{(n-1)})a(\tilde{u}^{(n-1)}+\tilde{v}_l^{(n-1)}+4\eta)},\nonumber\\[4mm]
&&\bar{\cal{{B}}}^{(m_{n-2},m_{n-1})}(u)=\prod_{k=1}^{m_{n-2}}
\frac{e_1(\tilde{u}^{(n-2)}+\tilde{v}_k^{(n-2)})e_1(\tilde{u}^{(n-2)}-\tilde{v}_k^{(n-2)})}
{e_{2}(\tilde{u}^{(n-2)}+\tilde{v}_k^{(n-2)})e_{2}(\tilde{u}^{(n-2)}-\tilde{v}_k^{(n-2)})}\nonumber\\
&&\hspace{30mm}\times\prod_{l=1}^{m_{n-1}}
\frac{a(\tilde{u}^{(n-1)}-\tilde{v}_l^{(n-1)})a(\tilde{u}^{(n-1)}+\tilde{v}_l^{(n-1)}-4\eta)}
{b(\tilde{u}^{(n-1)}-\tilde{v}_l^{(n-1)})a(\tilde{u}^{(n-1)}+\tilde{v}_l^{(n-1)}+4\eta)}.\nonumber
\end{eqnarray}
for $C^{(1)}_{n}$ and
\begin{eqnarray}
&&{\cal{B}}^{(m_{n-3},m_{n-2},m_{n-1})}(u)=\prod_{s=1}^{m_{n-3}}
\frac{a_2(\tilde{u}^{(n-3)}+\tilde{v}_s^{(n-3)})a_2(\tilde{u}^{(n-3)}-\tilde{v}_s^{(n-3)})}
{e_{3}(\tilde{u}^{(n-3)}+\tilde{v}_s^{(n-3)})e_{3}(\tilde{u}^{(n-3)}-\tilde{v}_s^{(n-3)})}\nonumber\\
&&\hspace{30mm}\times\prod_{k=1}^{m_{n-2}}
\frac{a(-\tilde{u}^{(n-2)}+\tilde{v}_k^{(n-2)})a(-\tilde{u}^{(n-2)}-\tilde{v}_k^{(n-2)})}
{b(-\tilde{u}^{(n-2)}+\tilde{v}_k^{(n-2)})b(-\tilde{u}^{(n-2)}-\tilde{v}_k^{(n-2)})}\nonumber\\
&&\hspace{30mm}\times\prod_{l=1}^{m_{n-1}}
\frac{a(-\tilde{u}^{(n-2)}+\tilde{v}_l^{(n-1)})a(-\tilde{u}^{(n-2)}-\tilde{v}_l^{(n-1)})}
{b(-\tilde{u}^{(n-2)}+\tilde{v}_l^{(n-1)})b(-\tilde{u}^{(n-2)}-\tilde{v}_l^{(n-1)})},\nonumber\\[4mm]
&&{\cal{B}}^{(m_{n-2},m_{n-1})}(u)=\prod_{k=1}^{m_{n-2}}
\frac{a(-\tilde{u}^{(n-2)}+\tilde{v}_k^{(n-2)})a(-\tilde{u}^{(n-2)}-\tilde{v}_k^{(n-2)})}
{b(-\tilde{u}^{(n-2)}+\tilde{v}_k^{(n-2)})b(-\tilde{u}^{(n-2)}-\tilde{v}_k^{(n-2)})}\nonumber\\
&&\hspace{30mm}\times\prod_{l=1}^{m_{n-1}}
\frac{a(\tilde{u}^{(n-2)}-\tilde{v}_l^{(n-1)})a(\tilde{u}^{(n-2)}+\tilde{v}_l^{(n-1)})}
{b(\tilde{u}^{(n-2)}-\tilde{v}_l^{(n-1)})b(\tilde{u}^{(n-2)}+\tilde{v}_l^{(n-1)})},\nonumber\\[4mm]
&&{\bar{\cal{B}}}^{(m_{n-3},m_{n-2},m_{n-1})}(u)=\prod_{s=1}^{m_{n-3}}
\frac{e_2(\tilde{u}^{(n-3)}+\tilde{v}_s^{(n-3)})e_2(\tilde{u}^{(n-3)}-\tilde{v}_s^{(n-3)})}
{e_{3}(\tilde{u}^{(n-3)}+\tilde{v}_s^{(n-3)})e_{3}(\tilde{u}^{(n-3)}-\tilde{v}_s^{(n-3)})}\nonumber\\
&&\hspace{30mm}\times\prod_{k=1}^{m_{n-2}}
\frac{a(\tilde{u}^{(n-2)}+\tilde{v}_k^{(n-2)})a(\tilde{u}^{(n-2)}-\tilde{v}_k^{(n-2)})}
{b(\tilde{u}^{(n-2)}+\tilde{v}_k^{(n-2)})b(\tilde{u}^{(n-2)}-\tilde{v}_k^{(n-2)})}\nonumber\\
&&\hspace{30mm}\times\prod_{l=1}^{m_{n-1}}
\frac{a(\tilde{u}^{(n-2)}+\tilde{v}_l^{(n-1)})a(\tilde{u}^{(n-2)}-\tilde{v}_l^{(n-1)})}
{b(\tilde{u}^{(n-2)}+\tilde{v}_l^{(n-1)})b(\tilde{u}^{(n-2)}-\tilde{v}_l^{(n-1)})},\nonumber\\[4mm]
&&{\bar{\cal{B}}}^{(m_{n-2},m_{n-1})}(u)=\prod_{k=1}^{m_{n-2}}
\frac{a(\tilde{u}^{(n-2)}+\tilde{v}_k^{(n-2)})a(\tilde{u}^{(n-2)}-\tilde{v}_k^{(n-2)})}
{b(\tilde{u}^{(n-2)}+\tilde{v}_k^{(n-2)})b(\tilde{u}^{(n-2)}-\tilde{v}_k^{(n-2)})}\nonumber\\
&&\hspace{30mm}\times\prod_{l=1}^{m_{n-1}}
\frac{a(-\tilde{u}^{(n-2)}-\tilde{v}_l^{(n-1)})a(-\tilde{u}^{(n-2)}+\tilde{v}_l^{(n-1)})}
{b(-\tilde{u}^{(n-2)}-\tilde{v}_l^{(n-1)})b(-\tilde{u}^{(n-2)}+\tilde{v}_l^{(n-1)})},\nonumber\\[4mm]
\end{eqnarray}
for $D^{(1)}_{n}$, where  $a(u)=\sinh(\frac{u}{2}-4\eta),
b(u)=\sinh(\frac{u}{2})$ for $C^{(1)}_{n}$  and
$a(u)=\sinh(\frac{u}{2}-2\eta), b(u)=\sinh(\frac{u}{2})$ for
$D^{(1)}_{n}$. The other Bethe equations are

\begin{eqnarray}
\displaystyle&&\prod_{k=1}^{m_{n-3}}\frac{a_{2}(\tilde{v}^{(n-2)}_i+\tilde{v}^{(n-3)}_k)
a_{2}(\tilde{v}^{(n-2)}_i-\tilde{v}^{(n-3)}_k)}
{b_{2}(\tilde{v}^{(n-2)}_i+\tilde{v}^{(n-3)}_k)b_{2}(\tilde{v}^{(n-2)}_i-\tilde{v}^{(n-3)}_k)}\nonumber\\
&&\times\prod_{l=1}^{m_{n-1}}\frac{e_1(-\tilde{v}^{(n-2)}_i+\tilde{v}^{(n-1)}_l)
e_1(-\tilde{v}^{(n-2)}_i-\tilde{v}^{(n-1)}_l)}
{a_1(-\tilde{v}^{(n-2)}_i+\tilde{v}^{(n-1)}_l)a_1(-\tilde{v}^{(n-2)}_i-\tilde{v}^{(n-1)}_l)}\nonumber\\
&&\times\prod_{s=1\ne i
}^{m_{n-2}}\frac{a_{2}(-\tilde{v}^{(n-2)}_i+\tilde{v}^{(n-2)}_s)
a_{2}(-\tilde{v}^{(n-2)}_i-\tilde{v}^{(n-2)}_s)}
{b_{2}(-\tilde{v}^{(n-2)}_i+\tilde{v}^{(n-2)}_s)b_{2}(-\tilde{v}^{(n-2)}_i-\tilde{v}^{(n-2)}_s)}
\frac{e_{2}(\tilde{v}^{(n-2)}_i+\tilde{v}^{(n-2)}_s)
e_{2}(\tilde{v}^{(n-2)}_i-\tilde{v}^{(n-2)}_s)}
{a_1(\tilde{v}^{(n-2)}_i+\tilde{v}^{(n-2)}_s)a_1(\tilde{v}^{(n-2)}_i-\tilde{v}^{(n-2)}_s)}\nonumber\\
&&=-\frac{{w}_1^{(n-1)}(\tilde{v}_i^{(n-2)})a_1(2\tilde{v}_i^{(n-2)}))
}{\beta_{n-1}(v_i^{(n-2)})}\frac{\bar{\omega}^{(n-2)}(v_i^{(n-2)})\bar{\omega}_1^{(n-1)}(\tilde{v}_i^{(n-2)})}
{\bar{\omega}_1^{(n-2)}({v}_i^{(n-2)})},
\hspace{2mm}(i=1,\cdots,m_{n-2})
\end{eqnarray}

\begin{eqnarray}
\displaystyle&&\prod_{k=1}^{m_{n-2}}\frac{a_1(\tilde{v}^{(n-1)}_i+\tilde{v}^{(n-2)}_k)
a_1(\tilde{v}^{(n-1)}_i-\tilde{v}^{(n-2)}_k)}
{e_1(\tilde{v}^{(n-1)}_i+\tilde{v}^{(n-2)}_k)e_1(\tilde{v}^{(n-1)}_i-\tilde{v}^{(n-2)}_k)}\nonumber\\
&&\times\prod_{l=1\ne
i}^{m_{n-1}}\frac{a_1(-\tilde{v}^{(n-1)}_i+\tilde{v}^{(n-1)}_l)
a_1(-\tilde{v}^{(n-1)}_i-\tilde{v}^{(n-1)}_l)}
{a_1(\tilde{v}^{(n-1)}_i-\tilde{v}^{(n-1)}_l)a_1(\tilde{v}^{(n-1)}_i+\tilde{v}^{(n-1)}_l)}\nonumber\\
&&=-\frac{\bar{\omega}_2^{(n-1)}(v_i^{(n-1)})}{\bar{\omega}_1^{(n-1)}(v_i^{(n-1)})\beta_{n}(v_i^{(n-1)})}
\hspace{2mm}(i=1,\cdots,m_{n-1})
\end{eqnarray}
for $A^{(2)}_{2n-1}$,
\begin{eqnarray}
\displaystyle&&\prod_{k=1}^{m_{n-3}}\frac{a_{2}(\tilde{v}^{(n-2)}_i+\tilde{v}^{(n-3)}_k)
a_{2}(\tilde{v}^{(n-2)}_i-\tilde{v}^{(n-3)}_k)}
{b_{2}(\tilde{v}^{(n-2)}_i+\tilde{v}^{(n-3)}_k)b_{2}(\tilde{v}^{(n-2)}_i-\tilde{v}^{(n-3)}_k)}\nonumber\\
&&\times\prod_{l=1}^{m_{n-1}}\frac{b(-\tilde{v}^{(n-2)}_i+\tilde{v}^{(n-1)}_l)
a(\tilde{v}^{(n-2)}_i+\tilde{v}^{(n-1)}_l+4\eta)}
{a(-\tilde{v}^{(n-2)}_i+\tilde{v}^{(n-1)}_l)b(\tilde{v}^{(n-2)}_i+\tilde{v}^{(n-1)}_l+4\eta)}\nonumber\\
&&\times\prod_{s=1\ne i
}^{m_{n-2}}\frac{a_{2}(-\tilde{v}^{(n-2)}_i+\tilde{v}^{(n-2)}_s)
a_{2}(-\tilde{v}^{(n-2)}_i-\tilde{v}^{(n-2)}_s)}
{b_{2}(-\tilde{v}^{(n-2)}_i+\tilde{v}^{(n-2)}_s)b_{2}(-\tilde{v}^{(n-2)}_i-\tilde{v}^{(n-2)}_s)}
\frac{e_{2}(\tilde{v}^{(n-2)}_i+\tilde{v}^{(n-2)}_s)
e_{2}(\tilde{v}^{(n-2)}_i-\tilde{v}^{(n-2)}_s)}
{a_1(\tilde{v}^{(n-2)}_i+\tilde{v}^{(n-2)}_s)a_1(\tilde{v}^{(n-2)}_i-\tilde{v}^{(n-2)}_s)}\nonumber\\
&&=-\frac{{w}_1^{(n-1)}(\tilde{v}_i^{(n-2)})a_1(2\tilde{v}_i^{(n-2)}))
}{\beta_{n-1}(v_i^{(n-2)})}\frac{\bar{\omega}^{(n-2)}(v_i^{(n-2)})\bar{\omega}_1^{(n-1)}(\tilde{v}_i^{(n-2)})}
{\bar{\omega}_1^{(n-2)}({v}_i^{(n-2)})},
\hspace{2mm}(i=1,\cdots,m_{n-2})\label{nestBAC1}
\end{eqnarray}
\begin{eqnarray}
\displaystyle&&\prod_{k=1}^{m_{n-2}}\frac{a_1(\tilde{v}^{(n-1)}_i+\tilde{v}^{(n-2)}_k)
a_1(\tilde{v}^{(n-1)}_i-\tilde{v}^{(n-2)}_k)}
{e_1(\tilde{v}^{(n-1)}_i+\tilde{v}^{(n-2)}_k)e_1(\tilde{v}^{(n-1)}_i-\tilde{v}^{(n-2)}_k)}\nonumber\\
&&\times\prod_{l=1\ne
i}^{m_{n-1}}\frac{a(-\tilde{v}^{(n-1)}_i+\tilde{v}^{(n-1)}_l)
b(-\tilde{v}^{(n-1)}_i-\tilde{v}^{(n-1)}_l-4\eta)}
{a(\tilde{v}^{(n-1)}_i-\tilde{v}^{(n-1)}_l)a(\tilde{v}^{(n-1)}_i+\tilde{v}^{(n-1)}_l-4\eta)}\nonumber\\
&&=-\frac{\bar{\omega}_2^{(n-1)}(v_i^{(n-1)})}{\bar{\omega}_1^{(n-1)}(v_i^{(n-1)})\beta_{n}(v_i^{(n-1)})}
\hspace{2mm}(i=1,\cdots,m_{n-1})\label{nestBAC2}
\end{eqnarray}
for $C^{(1)}_{n}$ and
\begin{eqnarray}
\displaystyle&&\prod_{k=1}^{m_{n-4}}\frac{a_{3}(\tilde{v}^{(n-3)}_i+\tilde{v}^{(n-4)}_k)
a_{3}(\tilde{v}^{(n-3)}_i-\tilde{v}^{(n-4)}_k)}
{b_{3}(\tilde{v}^{(n-3)}_i+\tilde{v}^{(n-4)}_k)b_{3}(\tilde{v}^{(n-3)}_i-\tilde{v}^{(n-4)}_k)}\nonumber\\
&&\times\prod_{l=1}^{m_{n-2}}\frac{b(-\tilde{v}^{(n-3)}_i+\tilde{v}^{(n-2)}_l)
b(-\tilde{v}^{(n-3)}_i-\tilde{v}^{(n-2)}_l)}
{a(-\tilde{v}^{(n-3)}_i+\tilde{v}^{(n-2)}_l)a(-\tilde{v}^{(n-3)}_i-\tilde{v}^{(n-2)}_l)}\nonumber\\
&&\times\prod_{t=1}^{m_{n-1}}\frac{b(-\tilde{v}^{(n-3)}_i+\tilde{v}^{(n-1)}_t)
b(-\tilde{v}^{(n-3)}_i-\tilde{v}^{(n-1)}_t)}
{a(-\tilde{v}^{(n-3)}_i+\tilde{v}^{(n-1)}_t)a(-\tilde{v}^{(n-3)}_i-\tilde{v}^{(n-1)}_t)}\nonumber\\
&&\times\prod_{s=1\ne i
}^{m_{n-3}}\frac{a_{3}(-\tilde{v}^{(n-3)}_i+\tilde{v}^{(n-3)}_s)
a_{3}(-\tilde{v}^{(n-3)}_i-\tilde{v}^{(n-3)}_s)}
{b_{3}(-\tilde{v}^{(n-3)}_i+\tilde{v}^{(n-3)}_s)b_{3}(-\tilde{v}^{(n-3)}_i-\tilde{v}^{(n-3)}_s)}
\frac{e_{3}(\tilde{v}^{(n-3)}_i+\tilde{v}^{(n-3)}_s)
e_{3}(\tilde{v}^{(n-3)}_i-\tilde{v}^{(n-3)}_s)}
{a_2(\tilde{v}^{(n-3)}_i+\tilde{v}^{(n-3)}_s)a_2(\tilde{v}^{(n-3)}_i-\tilde{v}^{(n-3)}_s)}\nonumber\\
&&=-\frac{{W}_1^1(\tilde{v}_i^{(n-3)}){W}_1^2(\tilde{v}_i^{(n-3)})a_2(2\tilde{v}_i^{(n-3)}))
}{\beta_{n-2}(v_i^{(n-3)})}
\frac{\bar{\omega}^{(n-3)}(v_i^{(n-3)}){\Omega}_1^1(\tilde{v}_i^{(n-3)}){\Omega}_1^2(\tilde{v}_i^{(n-3)})}
{\bar{\omega}_1^{(n-3)}({v}_i^{(n-3)})},\nonumber\\
&&\hspace{10mm}(i=1,\cdots,m_{n-3} )\label{nestBAD1}
\end{eqnarray}

\begin{eqnarray}
\displaystyle&&\prod_{k=1}^{m_{n-3}}\frac{a(\tilde{v}^{(n-2)}_i+\tilde{v}^{(n-3)}_k)
a(\tilde{v}^{(n-2)}_i-\tilde{v}^{(n-3)}_k)}
{b(\tilde{v}^{(n-2)}_i+\tilde{v}^{(n-3)}_k)b(\tilde{v}^{(n-2)}_i-\tilde{v}^{(n-3)}_k)}\nonumber\\
&&\times\prod_{l=1\ne
i}^{m_{n-2}}\frac{a(-\tilde{v}^{(n-2)}_i+\tilde{v}^{(n-2)}_l)
a(-\tilde{v}^{(n-2)}_i-\tilde{v}^{(n-2)}_l)}
{a(\tilde{v}^{(n-2)}_i-\tilde{v}^{(n-2)}_l)a(\tilde{v}^{(n-2)}_i+\tilde{v}^{(n-2)}_l)}\nonumber\\
&&=-\frac{{\Omega}_2^1(v_i^{(n-2)})}{{\Omega}_1^1({v}_i^{(n-2)})\beta_{n-1}(v_i^{(n-2)})}.
\hspace{2mm}(i=1,\cdots,m_{n-2})\label{nestBAD2}
\end{eqnarray}
\begin{eqnarray}
\displaystyle&&\prod_{k=1}^{m_{n-3}}\frac{a(\tilde{v}^{(n-1)}_i+\tilde{v}^{(n-3)}_k)
a(\tilde{v}^{(n-1)}_i-\tilde{v}^{(n-3)}_k)}
{b(\tilde{v}^{(n-1)}_i+\tilde{v}^{(n-3)}_k)b(\tilde{v}^{(n-1)}_i-\tilde{v}^{(n-3)}_k)}\nonumber\\
&&\times\prod_{l=1\ne
i}^{m_{n-1}}\frac{a(-\tilde{v}^{(n-1)}_i+\tilde{v}^{(n-1)}_l)
a(-\tilde{v}^{(n-1)}_i-\tilde{v}^{(n-1)}_l)}
{a(\tilde{v}^{(n-1)}_i-\tilde{v}^{(n-1)}_l)a(\tilde{v}^{(n-1)}_i+\tilde{v}^{(n-1)}_l)}\nonumber\\
&&=-\frac{{\Omega}_2^2(v_i^{(n-1)})}{{\Omega}_1^2({v}_i^{(n-1)})\beta_{n-1}(v_i^{(n-1)})}.
\hspace{2mm}(i=1,\cdots,m_{n-1})\label{nestBAD3}
\end{eqnarray}
for $D^{(1)}_n$.

The coefficients are

\begin{eqnarray}
&& \bar{\omega}_{1}^{(j)}(u^{(j)})=1,
\hspace{4mm}\bar{\omega}^{(j)}(u^{(j)})=\frac{e^{2\eta}\sinh(u^{(j)})}{\sinh(u^{(j)}-2\eta)},\nonumber
\\
&&
\bar{\omega}_{2(n-j)}^{(j)}(u^{(j)})=\frac{e^{2(2(n-j))\eta}\sinh(u^{(j)})\cosh(u^{(j)}-2((n-j)-1)\eta)}
{\sinh(u^{(j)}-2(2(n-j)-1)\eta)\cosh(u^{(j)}-2(n-j)\eta)}.
\end{eqnarray}
\begin{eqnarray}
&&
w_{1}^{(j)}(u^{(j)})=\frac{\sinh(u^{(j)}-2(2(n-j))\eta)\cosh(u^{(j)}-2((n-j)+1)\eta)}
{\sinh(u^{(j)}-2\eta)\cosh(u^{(j)}-2(n-j)\eta)},\nonumber\\
&&
w^{(j)}(u^{(j)})=\frac{e^{-2\eta}\sinh(u^{(j)}-2(2(n-j))\eta)}{\sinh(u^{(j)}-2(2(n-j)-1)\eta)},
\hspace{4mm}w_{2(n-j)}^{(j)}(u^{(j)})=e^{-2(2(n-j))\eta},
\end{eqnarray}
\begin{equation}
\beta_{j+1}(v_i^{(j)})=\displaystyle\left\{\begin{array}{ll}
\displaystyle-\frac{2e^{2\eta}\sinh(v_i^{(j)})\sinh(v_i^{(j)}-2(2(n-j)-1)\eta)
\cosh(v_i^{(j)}-2((n-j)+1)\eta)}{\sinh(v_i^{(j)}-2\eta)},&
j\le n-2\\
\displaystyle-\frac{e^{4\eta}\sinh(2v_i^{(n-1)})}{\sinh(2v_i^{(n-1)}-4\eta)}.&
j=n-1
\end{array}\right.
\end{equation}
for $A^{(2)}_{2n-1}$,

\begin{eqnarray}
&& \bar{\omega}_{1}^{(j)}(u^{(j)})=1,
\hspace{4mm}\bar{\omega}^{(j)}(u^{(j)})=\frac{e^{2\eta}\sinh(u^{(j)})}{\sinh(u^{(j)}-2\eta)},\nonumber
\\
&&
\bar{\omega}_{2(n-j)}^{(j)}(u^{(j)})=\frac{e^{2(2(n-j))\eta}\sinh(u^{(j)})\sinh(u^{(j)}-2((n-j)+2)\eta)}
{\sinh(u^{(j)}-2(2(n-j)+1)\eta)\sinh(u^{(j)}-2((n-j)+1)\eta)}.
\end{eqnarray}
\begin{eqnarray}
&&
w_{1}^{(j)}(u^{(j)})=\frac{\sinh(u^{(j)}-2(2(n-j)+2)\eta)\sinh(u^{(j)}-2(n-j)\eta)}
{\sinh(u^{(j)}-2\eta)\sinh(u^{(j)}-2((n-j)+1)\eta)},\nonumber\\
&&
w^{(j)}(u^{(j)})=\frac{e^{-2\eta}\sinh(u^{(j)}-2(2(n-j)+2)\eta)}{\sinh(u^{(j)}-2(2(n-j)+1)\eta)},
\hspace{4mm}w_{2(n-j)}^{(j)}(u^{(j)})=e^{-2(2(n-j))\eta},
\end{eqnarray}
\begin{equation}
\beta_{j+1}(v_i^{(j)})=\displaystyle\left\{\begin{array}{ll}
\displaystyle-\frac{2e^{2\eta}\sinh(v_i^{(j)})\sinh(v_i^{(j)}-2(2(n-j)+1)\eta)
\sinh(v_i^{(j)}-2(n-j)\eta)}{\sinh(v_i^{(j)}-2\eta)},&
j\le n-2\\
\displaystyle-\frac{e^{4\eta}\sinh(v_i^{(n-1)})}{\sinh(v_i^{(n-1)}-4\eta)}.&
j=n-1
\end{array}\right.
\end{equation}
for $C^{(1)}_n$ and

\begin{eqnarray}
&& \bar{\omega}_{1}^{(j)}(u^{(j)})=1,
\hspace{4mm}\bar{\omega}^{(j)}(u^{(j)})=\frac{e^{2\eta}\sinh(u^{(j)})}{\sinh(u^{(j)}-2\eta)},\nonumber
\\
&&
\bar{\omega}_{2(n-j)}^{(j)}(u^{(j)})=\frac{e^{2(2(n-j)-2)\eta}\sinh(u^{(j)})\sinh(u^{(j)}-2((n-j)-2)\eta)}
{\sinh(u^{(j)}-2(2(n-j)-3)\eta)\sinh(u^{(j)}-2((n-j)-1)\eta)}\nonumber\\
&& \hspace{11cm}(j\le n-3)
\end{eqnarray}
\begin{eqnarray}
&&
w_{1}^{(j)}(u^{(j)})=\frac{\sinh(u^{(j)}-2(2(n-j)-2)\eta)\sinh(u^{(j)}-2(n-j)\eta)}
{\sinh(u^{(j)}-2\eta)\sinh(u^{(j)}-2((n-j)-3)\eta)},\nonumber\\
&&
w^{(j)}(u^{(j)})=\frac{e^{-2\eta}\sinh(u^{(j)}-2(2(n-j)-2)\eta)}{\sinh(u^{(j)}-2(2(n-j)-3)\eta)},
\hspace{4mm}w_{2(n-j)}^{(j)}(u^{(j)})=e^{-2(2(n-j)-2)\eta},\nonumber\\
&& \hspace{11cm}(j\le n-3)
\end{eqnarray}
\begin{eqnarray}
&&{w}_1^{(n-2)}(u^{(n-2)})\bar{\omega}_1^{(n-2)}(u^{(n-2)})=
2W_1^1(u^{(n-2)})W_1^2(u^{(n-2)})\Omega_1^1(u^{(n-2)})\Omega_1^2(u^{(n-2)}),\nonumber\\
&&{w}_1^{(n-1)}(u^{(n-2)})\bar{\omega}_1^{(n-1)}(u^{(n-2)})=
2W_2^1(u^{(n-2)})W^2_1(u^{(n-2)})\Omega_2^1(u^{(n-2)})\Omega_1^2(u^{(n-2)}),\nonumber\\
&&{w}_4^{(n-2)}(u^{(n-2)})\bar{\omega}_4^{(n-2)}(u^{(n-2)})=
2W_1^1(u^{(n-2)})W^2_2(u^{(n-2)})\Omega_1^1(u^{(n-2)})\Omega^2_2(u^{(n-2)}),\nonumber\\
&&{w}_2^{(n-1)}(u^{(n-2)})\bar{\omega}_2^{(n-1)}(u^{(n-2)})=
2W_2^1(u^{(n-2)})W_2^2(u^{(n-2)})\Omega_2^1(u^{(n-2)})\Omega^2_2(u^{(n-2)})\nonumber
\end{eqnarray}
with
\begin{eqnarray}
\Omega_1^1(u)=\Omega^2_1(u)=1,&&\Omega_2^1(u)=\Omega^2_2(u)=\frac{e^{2\eta}\sinh(u)}{\sinh(u-2\eta)},\nonumber\\
W_1^1(u)=W^2_1(u)=\frac{\sinh(u-4\eta)}{\sinh(u-2\eta)},&&
W_2^1(u)=W^2_2(u)=e^{-2\eta}.
\end{eqnarray}

\begin{equation}
\beta_{j+1}(v_i^{(j)})=\displaystyle\left\{\begin{array}{l}
\displaystyle-\frac{2e^{2\eta}\sinh(v_i^{(j)})\sinh(v_i^{(j)}-2(2(n-j)-3)\eta)
\sinh(v_i^{(j)}-2(n-j)\eta)}{\sinh(v_i^{(j)}-2\eta)},\hspace{2mm}
j\le n-3\\
\displaystyle-\frac{e^{2\eta}\sinh(v_i^{(j)})}{\sinh(v_i^{(j)}-2\eta)}.\hspace{4cm}
j=n-2,n-1
\end{array}\right.
\end{equation}
for $D^{(1)}_n$. Up to now, we have gotten the whole eigenvalues
and the Bethe equations of transfer matrix   for the
$A^{(2)}_{2n},A^{(2)}_{2n-1},B^{(1)}_n,C^{(1)}_n,D^{(1)}_{n}$
vertex model with trivial K matrix. We have checked that our
results can agree with that obtained by the analytical Bethe
ansatz\cite{a2n2}. For the non-trivial K matrices, we still can
repeat the similar procedure as we have done above. In the
following section we will present some results for the non-trivial
cases.

\subsection{ Results for $B^{(1)}_n,C^{(1)}_n,D^{(1)}_{n}$ models with
  diagonal K-matrices (one free parameter case)}

In this section we will consider the
$B^{(1)}_n,C^{(1)}_n,D^{(1)}_{n}$ models with K matrices having
one free parameter. We do not discuss other non-trivial cases for
the sake of simplicity. We can find the expressions of eigenvalues
Eqs.(\ref{nesttu11},\ref{nesttu12},\ref{nesttu13}) and Bethe
equations
Eqs.(\ref{nestBAB1},\ref{nestBAB2},\ref{nestBAC1}-\ref{nestBAD3})
still hold for $B^{(1)}_n,C^{(1)}_n,D^{(1)}_{n}$ models. So we
only present the coefficients for these models.

For $B^{(1)}_n$, from Eq.(\ref{kn3bn}), we take
\begin{equation}
K^{(3)}_{+}({u},n,{\zeta}_{+})={K^{(3)}_{-}}(-u-\vartheta,n,{\zeta}_{+})^tM{(n)}.\label{knp3}
\end{equation}
Then we can obtain the coefficients
\begin{eqnarray}
&&
\bar{\omega}_{1}^{(0)}(u^{(0)})=e^{-u}\sinh(\zeta_-+\frac{u}{2})\sinh(\zeta_-+\frac{u}{2}-(2n-3)\eta)\nonumber\\
&&\bar{\omega}^{(0)}(u^{(0)})=\frac{\sinh(u)\sinh(\zeta_--\frac{u}{2}+2\eta)
\sinh(\zeta_-+\frac{u}{2}-(2n-3)\eta)}{\sinh(u-2\eta)},
\\
&&
\bar{\omega}_{2n+1}^{(0)}(u^{(0)})=e^{u}\sinh(\zeta_--\frac{u}{2})\sinh(\zeta_--\frac{u}{2}-(2n-3)\eta)\nonumber\\
&&
+\frac{2e^{u}\sinh((2n-1)\eta)\sinh(2\eta)\sinh(\zeta_-+\frac{u}{2})
\sinh(\zeta_-+\frac{u}{2}-(2n-3)\eta)}{\sinh(u-(2n-1)\eta)\sinh(u-4(n-1)\eta)}\nonumber\\
&&+\frac{e^{u}\sinh((2n-1)\eta)\cosh((2n-3)\eta)\sinh(\zeta_--\frac{u}{2})
\sinh(\zeta_-+\frac{u}{2}-(2n-3)\eta)}{\sinh(u-4(n-1)\eta)}.\nonumber
\end{eqnarray}
\begin{eqnarray}
&&
w_{1}^{(0)}(u^{(0)})=\frac{e^{u}\sinh(u-(2n+1)\eta)\sinh(u-2(2n-1)\eta)\sinh(\zeta_+-\frac{u}{2})
\sinh(\zeta_+-\frac{u}{2}-(2n-3)\eta)}{\sinh(u-2\eta)\sinh(u-(2n-1)\eta)},\nonumber\\
&&
w^{(0)}(u^{(0)})=\frac{\sinh(u-2(2n-1)\eta)\sinh(\zeta_+-\frac{u}{2})
\sinh(\zeta_++\frac{u}{2}-(2n-1)\eta)}{\sinh(u-4(n-1)\eta)},\\
&&w_{2n+1}^{(0)}(u^{(0)})=e^{-u}\sinh(\zeta_++\frac{u}{2}-(2n-1)\eta)
\sinh(\zeta_++\frac{u}{2}-4(n-1)\eta),
\end{eqnarray}
\begin{equation}
\beta_{1}(v_i^{(0)})=
\displaystyle-\frac{2e^{v_i}\sinh(v_i)\sinh(v_i-4(n-1)\eta)\sinh(v_i-(2n+1)\eta)
\sinh(\zeta_+-\frac{v_i}{2}-(2n-3)\eta)}{\sinh(v_i-2\eta)\sinh(\zeta_++\frac{v_i}{2}-(2n-1)\eta)}.
\end{equation}
While for $1\le j\le n-1 $, Eqs.(\ref{bnow}-\ref{bnbeta}) still
hold.

Similarly,  for $C^{(1)}_n$, by Eqs.(\ref{kn3cn}) and
(\ref{knp3}), we have
\begin{eqnarray}
&&
\bar{\omega}_{1}^{(j)}(u^{(j)})=e^{-\frac{u^{(j)}}{2}}\sinh(\zeta_-^{(j)}+\frac{u^{(j)}}{2}),
\hspace{4mm}\bar{\omega}^{(j)}(u^{(j)})=\frac{e^{\eta}\sinh(u^{(j)})}{\sinh(u^{(j)}-2\eta)},
\\
&&
\bar{\omega}_{2(n-j)}^{(j)}(u^{(j)})=\frac{e^{2(2(n-j)-1)\eta+\frac{u^{(j)}}{2}}
\sinh(u^{(j)})\sinh(u^{(j)}-2((n-j)+2)\eta)}
{\sinh(u^{(j)}-2(2(n-j)+1)\eta)\sinh(u^{(j)}-2((n-j)+1)\eta)}\nonumber\\
&& \hspace{3cm}\times
\sinh(\zeta_-^{(j)}-\frac{u^{(j)}}{2}+2((n-j)+1)\eta).\hspace{6mm}
j\le n-1\nonumber
\end{eqnarray}
\begin{eqnarray}
&&
w_{1}^{(j)}(u^{(j)})=\frac{e^{\frac{u^{(j)}}{2}}\sinh(\zeta_+^{(j)}-\frac{u^{(j)}}{2})
\sinh(u^{(j)}-2(2(n-j)+2)\eta)\sinh(u^{(j)}-2(n-j)\eta)}
{\sinh(u^{(j)}-2\eta)\sinh(u^{(j)}-2((n-j)+1)\eta)},\nonumber\\
&&
w^{(j)}(u^{(j)})=\frac{e^{-\eta}\sinh(u^{(j)}-2(2(n-j)+2)\eta)}{\sinh(u^{(j)}-2(2(n-j)+1)\eta)},\hspace{3cm}
j\le n-1\nonumber\\
&&w_{2(n-j)}^{(j)}(u^{(j)})=e^{-\frac{u^{(j)}}{2}-2((n-j)-1)\eta}\sinh(\zeta_+^{(j)}+\frac{u^{(j)}}{2}-2((n-j)+1)\eta),
\end{eqnarray}
\begin{equation}
\beta_{j+1}(v_i^{(j)})=\displaystyle\left\{\begin{array}{ll}
\displaystyle-\frac{2e^{2v_i^{(j)}+\eta}\sinh(v_i^{(j)})\sinh(v_i^{(j)}-2(2(n-j)+1)\eta)
}{\sinh(v_i^{(j)}-2\eta)}\\ \hspace{10mm}\times
\sinh(v_i^{(j)}-2(n-j)\eta)\sinh(\zeta_+^{(j)}-\frac{v_i^{(j)}}{2})
,&
j\le n-2\\
\displaystyle-\frac{e^{v_i^{(n-1)}}\sinh(v_i^{(n-1)})\sinh(\zeta_+^{(n-1)}-\frac{v_i^{(n-1)}}{2})}
{\sinh(v_i^{(n-1)}-4\eta)\sinh(\zeta_+^{(n-1)}+\frac{v_i^{(n-1)}}{2}-4\eta)},&
j=n-1
\end{array}\right.
\end{equation}
where $\zeta_{\mp}^{(j)}=\zeta_{\mp}\pm j\eta$

For $D^{(1)}_n$, there are seven 1-parameter K matrices. For the
sake of simplicity, we only consider one K matrix. We choose
Eq.(\ref{kn37}) and  the corresponding
$K^{(3)}_{+}({u},n,{\zeta}_{+})$ by Eq.(\ref{knp3}). Then we get
\begin{eqnarray}
&&
\bar{\omega}_{1}^{(j)}(u^{(j)})=e^{-u^{(j)}}\sinh(\zeta_-^{(j)}+\frac{u^{(j)}}{2})
\sinh(\zeta_-^{(j)}-\frac{u^{(j)}}{2}+(2(n-j)-4)\eta),\nonumber\\
&&\bar{\omega}^{(j)}(u^{(j)})=\frac{\sinh(u^{(j)})}{\sinh(u^{(j)}-2\eta)},
\\
&& \bar{\omega}_{2(n-j)}^{(j)}(u^{(j)})=\frac{e^{u^{(j)}}
\sinh(u^{(j)})\sinh(u^{(j)}-2((n-j)-2)\eta)}
{\sinh(u^{(j)}-2(2(n-j)-3)\eta)\sinh(u^{(j)}-2((n-j)-1)\eta)}\nonumber\\
&&\hspace{10mm}\times\sinh(\zeta_-^{(j)}+\frac{u^{(j)}}{2}-2\eta)
\sinh(\zeta_-^{(j)}-\frac{u^{(j)}}{2}+(2(n-j)-1)\eta)
\hspace{4mm}(j\le n-3)\nonumber
\end{eqnarray}
\begin{eqnarray}
&& w_{1}^{(j)}(u^{(j)})=\frac{e^{u^{(j)}}
\sinh(u^{(j)}-2(n-j)\eta)\sinh(u^{(j)}-4((n-j)-1)\eta)}
{\sinh(u^{(j)}-2\eta)\sinh(u^{(j)}-2((n-j)-1)\eta)}\nonumber\\
&&\hspace{20mm}\times\sinh(\zeta_+^{(j)}+\frac{u^{(j)}}{2})
\sinh(\zeta_+^{(j)}-\frac{u^{(j)}}{2}+(2(n-j)-4)\eta),\nonumber \\
&&
w^{(j)}(u^{(j)})=\frac{\sinh(u^{(j)}-4((n-j)-1)\eta)}{\sinh(u^{(j)}-2(2(n-j)-3)\eta)},\\
&&w_{2(n-j)}^{(j)}(u^{(j)})=e^{-u^{(j)}}\sinh(\zeta_+^{(j)}+\frac{u^{(j)}}{2}-2\eta)
\sinh(\zeta_+^{(j)}-\frac{u^{(j)}}{2}+2((n-j)-1)\eta),\hspace{4mm}(j\le
n-3)\nonumber
\end{eqnarray}

\begin{eqnarray}
&&\displaystyle\Omega_1^1(u^{(n-2)})=e^{-\frac{u^{(n-2)}}{2}}\sinh(\zeta_-^{(n-2)}+\frac{u^{(n-2)}}{2}),\nonumber\\
&&\displaystyle\Omega_2^1(u^{(n-2)})=\frac{e^{\frac{u^{(n-2)}}{2}}\sinh(u^{(n-2)})
\sinh(\zeta_-^{(n-2)}-\frac{u^{(n-2)}}{2}+2\eta)}
{\sinh(u^{(n-2)}-2\eta)},\nonumber\\
&&\displaystyle
W_1^1(u^{(n-2)})=\frac{e^{\frac{u^{(n-2)}}{2}}\sinh(u^{(n-2)}-4\eta)
\sinh(\zeta_+^{(n-2)}-\frac{u^{(n-2)}}{2})}
{\sinh(u^{(n-2)}-2\eta)},\nonumber\\
&&\displaystyle W_2^1(u^{(n-2)})=e^{-\frac{u^{(n-2)}}{2}}\sinh(\zeta_+^{(n-2)}+\frac{u^{(n-2)}}{2}-2\eta),\nonumber\\
&&\displaystyle\Omega_1^2(u^{(n-2)})=e^{-\frac{u^{(n-2)}}{2}}\sinh(\zeta_-^{(n-2)}-\frac{u^{(n-2)}}{2}),\nonumber\\
&&\displaystyle\Omega_2^2(u^{(n-2)})=\frac{e^{\frac{u^{(n-2)}}{2}}\sinh(u^{(n-2)})
\sinh(\zeta_-^{(n-2)}+\frac{u^{(n-2)}}{2}-2\eta)}
{\sinh(u^{(n-2)}-2\eta)},\nonumber\\
&&\displaystyle
W_1^2(u^{(n-2)})=\frac{e^{\frac{u^{(n-2)}}{2}}\sinh(u^{(n-2)}-4\eta)
\sinh(\zeta_+^{(n-2)}+\frac{u^{(n-2)}}{2})}
{\sinh(u^{(n-2)}-2\eta)},\nonumber\\
&&\displaystyle
W_2^2(u^{(n-2)})=e^{-\frac{u^{(n-2)}}{2}}\sinh(\zeta_+^{(n-2)}-\frac{u^{(n-2)}}{2}+2\eta),
\end{eqnarray}
\begin{equation}
\beta_{j+1}(v_i^{(j)})=\displaystyle\left\{\begin{array}{l}
\displaystyle-\frac{2e^{2v_i^{(j)}}\sinh(v_i^{(j)})\sinh(v_i^{(j)}-2(2(n-j)-3)\eta)
\sinh(v_i^{(j)}-2(n-j)\eta)}{\sinh(v_i^{(j)}-2\eta)}\\
\hspace{1cm}\times \sinh(\zeta_+^{(j)}+\frac{v_i^{(j)}}{2})
\sinh(\zeta_+^{(j)}-\frac{v_i^{(j)}}{2}+(2(n-j)-4)\eta),\hspace{2mm}
j\le n-3\\
\displaystyle-\frac{e^{v_i^{(n-2)}}\sinh(v_i^{(n-2)})\sinh(\zeta_+^{(n-2)}-\frac{v_i^{(n-2)}}{2})}
{\sinh(v_i^{(n-2)}-2\eta)\sinh(\zeta_+^{(n-2)}+\frac{v_i^{(n-2)}}{2})-2\eta)}, \hspace{2mm}j=n-2\\
\displaystyle-\frac{e^{v_i^{(n-1)}}\sinh(v_i^{(n-1)})\sinh(\zeta_+^{(n-2)}+\frac{v_i^{(n-1)}}{2})}
{\sinh(v_i^{(n-1)}-2\eta)\sinh(\zeta_+^{(n-2)}-\frac{v_i^{(n-1)}}{2})+2\eta)},\hspace{2mm}j=n-1
\end{array}\right.
\end{equation}
where $\zeta_{\mp}^{(j)}=\zeta_{\mp}+ j\eta$. Now we present some
ABA solutions to open $B^{(1)}_n,C^{(1)}_n,D^{(1)}_{n}$ vertex
models with non-trivial diagonal K matrices(one free parameter
case). Repeating the same program demonstrated in
sections(2.2-2.4), we can obtain the ABA solutions for other
non-trivial cases.

\section{Conclusions}

In the framework  of algebraic Bethe ansatz, we solve the
$A^{(2)}_{2n},A^{(2)}_{2n-1},B^{(1)}_n,C^{(1)}_n,D^{(1)}_{n}$
vertex model with trivial K matrix and find that our results agree
with that obtained by analytical Bethe ansatz method \cite{a2n2}.
Some results are presented for the
$B^{(1)}_n,C^{(1)}_n,D^{(1)}_{n}$ with their diagonal K matrices
having one free parameter. Using these results we can study the
boundary effects etc.

Here some open vertex models, such as $D^{(2)}_n$,  are not
considered  due to their complicated R matrices. For these models,
we need to generalize the ABA further.   Additionally,  for the
ABA has been generalized to the spin chain with non-diagonal
reflecting matrices\cite{cao}-\cite{yangwl},  it is an interesting
problem on how to apply the method to other models with higher
rank algebras.

\section*{ Acknowledgements}

This work is supported by  Program for  NCET under Grant No.
NCET-04-0929 and NNSFC under Grant No.10575080.


\setcounter{section}{0}
\renewcommand{\thesection}{\Alph{section}}

\section{R matric elements}
\setcounter{equation}{0}
\renewcommand{\theequation}{A.\arabic{equation}}
The R matric elements are
\begin{eqnarray}
a_n(u)&=&2\sinh(\frac{u}{2}-2\eta)\sinh(\frac{u}{2}-\kappa\eta),\nonumber\\
b_n(u)&=&2\sinh(\frac{u}{2})\sinh(\frac{u}{2}-\kappa\eta),\nonumber\\
c_n(u,i)&=&d_n(u,i,\bar{i})-2e^{-\frac{u}{2}}\sinh(2\eta)\sinh(\frac{u}{2}-\kappa\eta),\nonumber\\
\bar{c}_n(u,i)&=&\bar{d}_n(u,i,\bar{i})-2e^{\frac{u}{2}}\sinh(2\eta)\sinh(\frac{u}{2}-\kappa\eta),\nonumber\\
d_n(u,i,j)&=&2\epsilon_i\epsilon_{j}e^{-\frac{u}{2}+(\kappa+2(\tilde{i}-\tilde{j}))\eta}\sinh(2\eta)\sinh(\frac{u}{2}),\nonumber\\
\bar{d}_n(u,i,j)&=&2\epsilon_i\epsilon_{j}e^{\frac{u}{2}+(2(\tilde{i}-\tilde{j})-\kappa)\eta}\sinh(2\eta)\sinh(\frac{u}{2}),\nonumber\\
e_n(u)&=&2\sinh(\frac{u}{2})\sinh(\frac{u}{2}-(\kappa-2)\eta),\nonumber\\
f_n(u)&=&b_n(u)+2\sinh(2\eta)\sinh(\kappa\eta),\nonumber\\
g_n(u)&=&-2e^{-\frac{u}{2}}\sinh(2\eta)\sinh(\frac{u}{2}-\kappa\eta),\nonumber\\
\bar{g}_n(u)&=&-2e^{\frac{u}{2}}\sinh(2\eta)\sinh(\frac{u}{2}-\kappa\eta)
 \label{2}
\end{eqnarray}
for $B^{(1)}_n, C^{(1)}_n, D^{(1)}_n$ and

\begin{eqnarray}
a_n(u)&=&2\sinh(\frac{u}{2}-2\eta)\cosh(\frac{u}{2}-\kappa\eta),\nonumber\\
b_n(u)&=&2\sinh(\frac{u}{2})\cosh(\frac{u}{2}-\kappa\eta),\nonumber\\
c_n(u,i)&=&d_n(u,i,\bar{i})-2e^{-\frac{u}{2}}\sinh(2\eta)\cosh(\frac{u}{2}-\kappa\eta),\nonumber\\
\bar{c}_n(u,i)&=&\bar{d}_n(u,i,\bar{i})-2e^{\frac{u}{2}}\sinh(2\eta)\cosh(\frac{u}{2}-\kappa\eta),\nonumber\\
d_n(u,i,j)&=&-2\epsilon_i\epsilon_{j}e^{-\frac{u}{2}+(\kappa+2(\tilde{i}-\tilde{j}))\eta}\sinh(2\eta)\sinh(\frac{u}{2}),\nonumber\\
\bar{d}_n(u,i,j)&=&2\epsilon_i\epsilon_{j}e^{\frac{u}{2}+(2(\tilde{i}-\tilde{j})-\kappa)\eta}\sinh(2\eta)\sinh(\frac{u}{2}),\nonumber\\
e_n(u)&=&2\sinh(\frac{u}{2})\cosh(\frac{u}{2}-(\kappa-2)\eta),\nonumber\\
f_n(u)&=&b_n(u)-2\sinh(2\eta)\cosh(\kappa\eta),\nonumber\\
g_n(u)&=&-2e^{-\frac{u}{2}}\sinh(2\eta)\cosh(\frac{u}{2}-\kappa\eta),\nonumber\\
\bar{g}_n(u)&=&-2e^{\frac{u}{2}}\sinh(2\eta)\cosh(\frac{u}{2}-\kappa\eta)
 \label{2}
\end{eqnarray}
for $A_{2n}^{(2)}, A_{2n-1}^{(2)}$. Where $\epsilon_i=1$ for
$A_{2n}^{(2)}, B^{(1)}_n, D^{(1)}_n$ models, $\epsilon_i=1$ if
$1\le i\le n$ and $\epsilon_i=-1$ if $n+1\le i\le 2n$ for
$A_{2n-1}^{(2)}, C^{(1)}_n$ models.

\section{ Derivations of diagonal operators acting on eigenvector}
\setcounter{equation}{0}
\renewcommand{\theequation}{B.\arabic{equation}}

Acting the diagonal operators
$x(u)=A(u),\tilde{A}_{aa}(u),\tilde{A}_2(u)$ on the m-particle
state and having carried out a very involved analysis as that in
Ref.\cite{ik}, we can obtain the following expression
\begin{eqnarray}
& &
\hspace{-10mm}{x}(u)|\Upsilon_m(v_1,\cdots,v_m)\rangle=|\Psi_{x}(u,\{v_m\})\rangle\nonumber\\
 & &+\sum_{i=1}^m h_1^x(u,v_i,d)|\Psi_{m-1}^{(1)}(u,v_i;\{v_m\})_{dd}\rangle\nonumber\\
& & +\sum_{i=1}^m  h_2^x(u,v_i,d)|\Psi_{m-1}^{(2)}(u,v_i;\{v_m\})_{dd}\rangle\nonumber\\
& &+\sum_{i=1}^m h_3^x(u,v_i,\bar{\alpha}_x)|\Psi_{m-1}^{(3)}(u,v_i;\{v_m\})_{\alpha_x\alpha_x}\rangle\nonumber\\
& &+\sum_{i=1}^m h_4^x(u,v_i,\bar{\alpha}_x)|\Psi_{m-1}^{(4)}(u,v_i;\{v_m\})_{\alpha_x\alpha_x}\rangle\nonumber\\
& & +\sum^{m-1}_{i=1}\sum_{j=i+1}^m \delta_{\bar{d}_1e_2}
H^{{x}}_{1,d_1}(u,v_i,v_j)|\Psi_{m-2}^{(5)}(u,v_i,v_j;\{v_m\})_{d_1e_2}\rangle\nonumber\\
& & +\sum^{m-1}_{i=1}\sum_{j=i+1}^m
H^{{x}}_{2,d_1}(u,v_i,v_j)^{fe}_{c_2d_1}|\Psi_{m-2}^{(6)}(u,v_i,v_j;\{v_m\})_{d_1c_2}^{ef}\rangle\nonumber\\
& & +\sum^{m-1}_{i=1}\sum_{j=i+1}^m H^{{x}}_{3,d_1}(u,v_i,v_j)
|\Psi_{m-2}^{(7)}(u,v_i,v_j;\{v_m\})_{d_1d_1}\rangle\nonumber\\
& & +\sum^{m-1}_{i=1}\sum_{j=i+1}^m
H^{{x}}_{4,d_1}(u,v_i,v_j)^{de}_{fd_1}|\Psi_{m-2}^{(8)}(u,v_i,v_j;\{v_m\})_{d_1f}^{ed}\rangle,
\label{d1pns}
\end{eqnarray}
where when $x=A,\tilde{A}_{aa},\tilde{A}_2$,
\begin{eqnarray}
&&h_1^x(u,v_i,d)=a^1_2(u,v_i),
R^A_1(u,v_i)^{da}_{ad},a^3_2(u,v_i),\nonumber\\
&&h_2^x(u,v_i,d)=a^1_3(u,v_i),
R^A_2(u,v_i)^{da}_{ad},a^3_3(u,v_i),\nonumber\\
&&h_3^x(u,v_i,\bar{\alpha}_x)=0,
R^A_3(u,v_i,\bar{a}),a^3_4(u,v_i,\bar{d}),\nonumber\\
&&h_4^x(u,v_i,\bar{\alpha}_x)=0,
R^A_4(u,v_i,\bar{a}),a^3_5(u,v_i,\bar{d}),\nonumber\\
&&\alpha_x=0, a, d, \nonumber
\end{eqnarray}
\begin{eqnarray}
|\Psi_{x}(u,\{v_m\})\rangle=\left\{\begin{array}{l}\omega_1(u)\Lambda_1^m(u;v_1,\cdots,v_m)
|\Upsilon_m(v_1,\cdots,v_m)\rangle,\\[2mm]
\Phi_m^{d_1\cdots
d_m}(v_1,\cdots,v_m)[\tilde{T}^{m}(u;\{{v}_m\})^{d_1\cdots
d_m}_{b_1\cdots b_m}]_{aa}F^{b_1\cdots b_m}|0\rangle,\\[2mm]
\omega_{q}(u)\Lambda_3^m(u;v_1,\cdots,v_m)
|\Upsilon_m(v_1,\cdots,v_m)\rangle, \end{array}\right.
\end{eqnarray}
respectively, and we denote
\begin{eqnarray}
& & \hspace{-6mm}|\Psi_{m-1}^{(1)}(u,v_i;\{v_m\})_{fd_1}\rangle=
B_f(u)|\Phi_{m-1}^{d_2\cdots
d_m}(v_1,\cdots,\check{v}_i,\cdots,v_m) \nonumber \\
& & \hspace{8mm}\times S^{d_1\cdots d_m}_{b_1\cdots
b_m}(v_i;\{\check{v}_i\})\Lambda_1^{m-1}(v_i; \{\check{v}_i\})
\omega_1(v_i)F^{b_1\cdots b_m}|0\rangle,\\[4mm]
& &
\hspace{-6mm}|\Psi_{m-1}^{(2)}(u,v_i;\{v_m\})_{fd}\rangle=B_f(u)\Phi_{m-1}^{e_2\cdots
e_m}(v_1,\cdots,\check{v}_i,\cdots,v_m)\nonumber\\
& & \hspace{10mm}
\times[\tilde{T}^{m-1}(v_i;\{\check{v}_i\})^{e_2\cdots
e_m}_{d_2'\cdots d_m'}]_{dd_1'}  S^{d_1'\cdots d_m'}_{b_1\cdots
b_m}(v_i;\{\check{v}_i\})F^{b_1\cdots b_m}|0\rangle,\\[4mm]
& & \hspace{-6mm}|\Psi_{m-1}^{(3)}(u,v_i;\{v_m\})_{ab}\rangle=
E_{{a}}(u)\Phi_{m-1}^{d_2\cdots
d_m}(v_1,\cdots,\check{v}_i,\cdots,v_m) \nonumber\\
& & \hspace{10mm} \times S^{d_1\cdots d_m}_{b_1\cdots
b_m}(v_i;\{\check{v}_i\})\Lambda_1^{m-1}(v_i; \{\check{v}_i\})
\omega_1(v_i) \delta_{\bar{b}d_1}F^{b_1\cdots b_m}|0\rangle,\\[4mm]
& & \hspace{-6mm}
|\Psi_{m-1}^{(4)}(u,v_i;\{v_m\})_{ab}\rangle=E_a(u)\Phi_{m-1}^{e_2\cdots
e_m}(v_1,\cdots,\check{v}_i,\cdots,v_m)\nonumber\\
& & \hspace{10mm}\times
[\tilde{T}^{m-1}(v_i;\{\check{v}_i\})^{e_2\cdots e_m}_{d_2'\cdots
d_m'}]_{\bar{b}d_1'} S^{d_1'\cdots d_m'}_{b_1\cdots
b_m}(v_i;\{\check{v}_i\}) F^{b_1\cdots b_m}|0\rangle,\\[4mm]
& &\hspace{-6mm}
|\Psi_{m-2}^{(5)}(u,v_i,v_j;\{v_m\})_{d_1e_2}\rangle =
F(u)\Phi_{m-2}^{e_3\cdots e_m}(v_1,\cdots,\check{v}_i,\cdots,\check{v}_j,\cdots,v_m)\nonumber \\
& & \hspace{4mm}\times S^{e_2\cdots e_m}_{d_2\cdots
d_m}(v_j;\{\check{v}_i,\check{v}_j\})S^{d_1\cdots d_m}_{b_1\cdots
b_m}(v_i;\{\check{v}_i\})\Lambda_1^{m-2}(v_i;
\{\check{v}_i,\check{v}_j\})\nonumber \\
& & \hspace{4mm}\times \Lambda_1^{m-2}(v_j;
\{\check{v}_i,\check{v}_j\})A(v_i)A(v_j)F^{b_1\cdots
b_m}|0\rangle,\\[4mm]
& &\hspace{-6mm}
|\Psi_{m-2}^{(6)}(u,v_i,v_j;\{v_m\})_{d_1c_2}^{ef}\rangle =
F(u)\Phi_{m-2}^{e_3\cdots e_m}(v_1,\cdots,\check{v}_i,\cdots,\check{v}_j,\cdots,v_m)\nonumber \\
& & \hspace{4mm}\times
[\tilde{T}^{m-2}(v_i;\{\check{v}_i,\check{v}_j\})^{e_3\cdots
e_m}_{c_3\cdots c_m}]_{\bar{f}e} S^{c_2\cdots c_m}_{d_2\cdots
d_m}(v_j;\{\check{v}_i,\check{v}_j\})S^{d_1\cdots d_m}_{b_1\cdots
b_m}(v_i;\{\check{v}_i\})\nonumber \\
& & \hspace{4mm}\times \Lambda_1^{m-2}(v_j;
\{\check{v}_i,\check{v}_j\})A(v_j)F^{b_1\cdots b_m}|0\rangle,\\[4mm]
& &\hspace{-6mm}
|\Psi_{m-2}^{(7)}(u,v_i,v_j;\{v_m\})_{fd_1}\rangle =
F(u)\Phi_{m-2}^{e_3\cdots e_m}(v_1,\cdots,\check{v}_i,\cdots,\check{v}_j,\cdots,v_m)\nonumber \\
& & \hspace{4mm}\times
[\tilde{T}^{m-2}(v_j;\{\check{v}_i,\check{v}_j\})^{e_3\cdots
e_m}_{c_3\cdots c_m}]_{\bar{f}c_2} S^{c_2\cdots c_m}_{d_2\cdots
d_m}(v_j;\{\check{v}_i,\check{v}_j\})S^{d_1\cdots d_m}_{b_1\cdots
b_m}(v_i;\{\check{v}_i\})\nonumber \\
& & \hspace{4mm}\times \Lambda_1^{m-2}(v_i;
\{\check{v}_i,\check{v}_j\})A(v_i)F^{b_1\cdots b_m}|0\rangle,\\[4mm]
& &\hspace{-6mm}
|\Psi_{m-2}^{(8)}(u,v_i,v_j;\{v_m\})_{d_1f}^{ed}\rangle =
F(u)\Phi_{m-2}^{e_3\cdots e_m}(v_1,\cdots,\check{v}_i,\cdots,\check{v}_j,\cdots,v_m)\nonumber \\
& & \hspace{4mm}\times
[\tilde{T}^{m-2}(v_i;\{\check{v}_i,\check{v}_j\})^{e_3\cdots
e_m}_{a_3\cdots a_m}]_{\bar{d}e}
[\tilde{T}^{m-2}(v_j;\{\check{v}_i,\check{v}_j\})^{a_3\cdots
a_m}_{c_3\cdots c_m}]_{fc_2}\nonumber \\
& & \hspace{4mm}\times S^{c_2\cdots c_m}_{d_2\cdots
d_m}(v_j;\{\check{v}_i,\check{v}_j\})S^{d_1\cdots d_m}_{b_1\cdots
b_m}(v_i;\{\check{v}_i\}) F^{b_1\cdots b_m}|0\rangle.
\end{eqnarray}
The explicit expressions of $H^{x}_{l,d_1}(u,v_i,v_j), l=1,2,3,4$
are listed as below
\begin{eqnarray}
 & & \hspace{-6mm} H^{A}_{1,d_1}(u,v_i,v_j)= a^1_4(u,v_i,\bar{d}_1)(c^1_5(v_i,v_j)+c^1_7(v_i,v_j))
  \nonumber\\
& &+
a^1_5(u,v_i)(c^2_4(v_i,v_j,\bar{d}_1)+c^2_6(v_i,v_j,\bar{d}_1))+b^1_2(u,v_i)g_{1}(v_i,v_j,d_1)\nonumber\\
& & +a^1_1(u,v_i)a^1_2(u,v_j)g_1(u,v_i,d)
\hat{r}(v_i-u)^{\bar{d}d}_{d_1\bar{d}_1},\\[4mm]
 & & \hspace{-6mm} H^{A}_{2,d_1}(u,v_i,v_j)^{fe}_{c_2d_1}=
 a^1_4(u,v_i,\bar{d}_1)(c^1_6(v_i,v_j)+c^1_9(v_i,v_j))\delta_{d_1f}
  \nonumber\\
& &+
a^1_5(u,v_i)(R^C_3(v_i,v_j)^{fe}_{c_2d_1}+R^C_4(v_i,v_j)^{fe}_{c_2d_1})\nonumber\\
& &+b^1_3(u,v_i)g_1(v_i,v_j,d_1)\delta_{d_1\bar{c}_2}+
a^1_1(u,v_i)a^1_2(u,v_j)g_2(u,v_i,f) \hat{r}(v_i-u)^{fe}_{c_2d_1},\\[4mm]
 & & \hspace{-6mm} H^{A}_{3,d_1}(u,v_i,v_j)= a^1_4(u,v_i,\bar{d}_1)c^1_8(v_i,v_j)
  + a^1_5(u,v_i)c^2_7(v_i,v_j,\bar{d}_1)\nonumber\\
&
&+b^1_2(u,v_i)g_{2}(v_i,v_j,d_1)+a^1_1(u,v_i)a^1_3(u,v_j)g_1(u,v_i,d)
\hat{r}(v_i-u)^{d\bar{d}}_{\bar{d}_1d_1},\\[4mm]
 & & \hspace{-6mm} H^{A}_{4,d_1}(u,v_i,v_j)^{de}_{fd_1}=
 a^1_4(u,v_i,\bar{d}_1)c^1_{10}(v_i,v_j)\delta_{d_1d}+
a^1_5(u,v_i)R^C_5(v_i,v_j)^{de}_{fd_1}\nonumber\\
& &+b^1_3(u,v_i)g_2(v_i,v_j,d_1)\delta_{\bar{d}_1f}+
a^1_1(u,v_i)a^1_3(u,v_j)g_2(u,v_i,d) \hat{r}(v_i-u)^{de}_{fd_1},\\[4mm]
 & & \hspace{-6mm} H^{\tilde{A}_{aa}}_{1,d_1}(u,v_i,v_j)= R^A_5(u,v_i)^{ad_1}_{d_1a}(c^1_5(v_i,v_j)+c^1_7(v_i,v_j))
  +b^2_2(u,v_i)g_{1}(v_i,v_j,d_1)\nonumber\\
& &+
R^A_6(u,v_i)(c^2_4(v_i,v_j,\bar{d}_1)+c^2_6(v_i,v_j,\bar{d}_1))
+\tilde{r}(u+v_i)^{ad}_{da}\bar{r}(u-v_i)^{ad_1}_{d_1a}R^A_3(u,v_j,\bar{d}_1)e^1_1(v_i,u,d)\nonumber\\
& &
+\tilde{r}(u+v_i)^{ae}_{dg}\bar{r}(u-v_i)^{gf}_{d_1a}g_1(u,v_i,h)R^A_{1}(u,v_j)^{d\bar{e}}_{\bar{d}_1f}
\hat{r}(v_i-u)^{h\bar{h}}_{\bar{e}e},\\[4mm]
 & & \hspace{-6mm} H^{\tilde{A}_{aa}}_{2,d_1}(u,v_i,v_j)^{fe}_{c_2d_1}=
 R^A_5(u,v_i)^{ad_1}_{d_1a}(c^1_6(v_i,v_j)+c^1_9(v_i,v_j))\delta_{d_1f}
  \nonumber\\
& &+
R^A_6(u,v_i)(R^C_3(v_i,v_j)^{fe}_{c_2d_1}+R^C_4(v_i,v_j)^{fe}_{c_2d_1})
+R^F_1(u,v_i)^{fe}_{a\bar{a}}g_1(v_i,v_j,d_1)\delta_{d_1\bar{c}_2}\nonumber\\
& &+
\tilde{r}(u+v_i)^{ac}_{dg}\bar{r}(u-v_i)^{gh}_{d_1a}g_2(u,v_i,f)R^A_{1}(u,v_j)^{db}_{c_2h}
\hat{r}(v_i-u)^{fe}_{bc}\nonumber\\
& &
+\tilde{r}(u+v_i)^{ah}_{dg}\bar{r}(u-v_i)^{g\bar{c}_2}_{d_1a}R^A_{3}(u,v_j,\bar{c}_2)
R^{be}_3(v_i,u)^{fe}_{\bar{d}h},\\[4mm]
 & & \hspace{-6mm} H^{\tilde{A}_{aa}}_{3,d_1}(u,v_i,v_j)= R^A_5(u,v_i)^{ad_1}_{d_1a}c^1_8(v_i,v_j)
  + R^A_6(u,v_i)c^2_7(v_i,v_j,\bar{d}_1)\nonumber\\
&
&+b^2_2(u,v_i)g_{2}(v_i,v_j,d_1)+\tilde{r}(u+v_i)^{ad}_{da}\bar{r}(u-v_i)^{ad_1}_{d_1a}R^A_4(u,v_j,\bar{d}_1)e^1_1(v_i,u,d)\nonumber\\
& &
+\tilde{r}(u+v_i)^{ae}_{dg}\bar{r}(u-v_i)^{gf}_{d_1a}g_1(u,v_i,h)R^A_{2}(u,v_j)^{d\bar{e}}_{\bar{d}_1f}
\hat{r}(v_i-u)^{h\bar{h}}_{\bar{e}e},\\[4mm]
 & & \hspace{-6mm} H^{\tilde{A}_{aa}}_{4,d_1}(u,v_i,v_j)^{de}_{fd_1}=
 R^A_5(u,v_i)^{ad_1}_{d_1a}c^1_{10}(v_i,v_j)\delta_{d_1d}+
R^A_6(u,v_i)R^C_5(v_i,v_j)^{de}_{fd_1}\nonumber\\
&
&+R^F_1(u,v_i)^{de}_{a\bar{a}}g_2(v_i,v_j,d_1)\delta_{d_1\bar{f}}
+\tilde{r}(u+v_i)^{ac}_{hg}\bar{r}(u-v_i)^{g\bar{f}}_{d_1a}R^A_{4}(u,v_j,f)
R^{be}_3(v_i,u)^{de}_{\bar{h}c}\nonumber\\
& & +
\tilde{r}(u+v_i)^{ac}_{bg}\bar{r}(u-v_i)^{gh}_{d_1a}g_2(u,v_i,d)R^A_{2}(u,v_j)^{bk}_{fh}
\hat{r}(v_i-u)^{de}_{kc},\\[4mm]
 & & \hspace{-6mm} H^{\tilde{A}_{2}}_{1,d_1}(u,v_i,v_j)= a^3_6(u,v_i,\bar{d}_1)(c^1_5(v_i,v_j)+c^1_7(v_i,v_j))
  +b^3_2(u,v_i)g_{1}(v_i,v_j,d_1)\nonumber\\
& &+
a^3_7(u,v_i)(c^2_4(v_i,v_j,\bar{d}_1)+c^2_6(v_i,v_j,\bar{d}_1))
+a^3_1(u,v_i)a^3_4(u,v_j,\bar{d}_1)e^1_1(v_i,u,d_1)\nonumber\\
& & +a^3_1(u,v_i)a^3_2(u,v_j)g_1(u,v_i,d)
\hat{r}(v_i-u)^{d\bar{d}}_{\bar{d}_1d_1},\\[4mm]
 & & \hspace{-6mm} H^{\tilde{A}_{2}}_{2,d_1}(u,v_i,v_j)^{fe}_{c_2d_1}=
 a^3_6(u,v_i,\bar{d}_1)(c^1_6(v_i,v_j)+c^1_9(v_i,v_j))\delta_{d_1f}
  \nonumber\\
& &+
a^3_7(u,v_i)(R^C_3(v_i,v_j)^{fe}_{c_2d_1}+R^C_4(v_i,v_j)^{fe}_{c_2d_1})\nonumber\\
&
&+b^3_3(u,v_i)g_1(v_i,v_j,d_1)\delta_{d_1\bar{c}_2}+a^3_1(u,v_i)a^3_4(u,v_j,c_2)
R^{be}_3(v_i,u)^{fe}_{c_2d_1}\nonumber\\
& & + a^3_1(u,v_i)a^3_2(u,v_j)g_2(u,v_i,f)
\hat{r}(v_i-u)^{fe}_{c_2d_1},\\[4mm]
 & & \hspace{-6mm} H^{\tilde{A}_{2}}_{3,d_1}(u,v_i,v_j)= a^3_6(u,v_i,\bar{d}_1)c^1_8(v_i,v_j)
  + a^3_7(u,v_i)c^2_7(v_i,v_j,\bar{d}_1)\nonumber\\
&
&+b^3_2(u,v_i)g_{2}(v_i,v_j,d_1)+a^3_1(u,v_i)a^3_5(u,v_j,d_1)e^1_1(v_i,u,d_1)\nonumber\\
& & +a^3_1(u,v_i)a^3_3(u,v_j)g_1(u,v_i,d)
\hat{r}(v_i-u)^{d\bar{d}}_{\bar{d}_1d_1},\\[4mm]
 & & \hspace{-6mm} H^{\tilde{A}_{2}}_{4,d_1}(u,v_i,v_j)^{de}_{fd_1}=
 a^3_6(u,v_i,\bar{d}_1)c^1_{10}(v_i,v_j)\delta_{d_1d}+
a^3_7(u,v_i)R^C_5(v_i,v_j)^{de}_{fd_1}\nonumber\\
& &+b^3_3(u,v_i)g_2(v_i,v_j,d_1)\delta_{d_1\bar{f}}
+a^3_1(u,v_i)a^3_5(u,v_j,\bar{f}) R^{be}_3(v_i,u)^{de}_{fd_1}\nonumber\\
& &+ a^3_1(u,v_i)a^3_3(u,v_j)g_2(u,v_i,d)
\hat{r}(v_i-u)^{de}_{fd_1}.
\end{eqnarray}
 In Eqs.(A.11)-(A.22),  the repeated indices  $a$ and $
d_1$ do not sum.  We can  check that
\begin{eqnarray}
\frac{H^{x}_{2,b}(u,v_i,v_j)^{{d}_1a}_{cb}}{R^{(n-1)}(\tilde{v}_i-\tilde{v}_j)^{{d}_1a}_{cb}}=
\frac{H^{x}_{2,{d}_1}(u,v_i,v_j)^{{d}_1{d}_1}_{{d}_1{d}_1}}{R^{(n-1)}(\tilde{v}_i-\tilde{v}_j)^{{d}_1{d}_1}_{{d}_1{d}_1}},
\label{haa2}
\end{eqnarray}
\begin{eqnarray}
\frac{H^{x}_{4,b}(u,v_i,v_j)^{{d}_1a}_{cb}}{R^{(n-1)}(\tilde{v}_i+\tilde{v}_j)^{{d}_1a}_{cb}}=
\frac{H^{x}_{4,{d}_1}(u,v_i,v_j)^{{d}_1{d}_1}_{{d}_1{d}_1}}{R^{(n-1)}(\tilde{v}_i+\tilde{v}_j)^{{d}_1{d}_1}_{{d}_1{d}_1}}.
\label{haa4}
\end{eqnarray}
In Eqs.(\ref{haa2},\ref{haa4}), all the repeated indices do not
sum. We conclude that  Eq.(\ref{d1pns}) can be verified directly
by using mathematical induction, although it is a rather hard
work. Similar to assumption of algebraic Bethe ansatz, we might
assume that ``quasi'' m-particle states such as
$B\Phi_{m-1}|0\rangle$, $E\Phi_{m-1}|0\rangle$,
$BB\Phi_{m-2}|0\rangle$, $BE\Phi_{m-2}|0\rangle$,
$EB\Phi_{m-2}|0\rangle$, $F\Phi_{m-2}|0\rangle$,
$FB\Phi_{m-3}|0\rangle$ etc are linearly independent. Here
 all the indices are omitted and all the spectrum parameters in the
``quasi'' n-particle state keep the order
$\{v_{i_1},v_{i_2},\cdots,v_{i_k}\}$ with\  $i_1 <i_2 < \cdots
<i_k$. For example, $B_{1}(v_1) B_{1}(v_2) \Phi_{m-2}^{b_3\cdots
b_m}$ $(v_3,\cdots,v_m) F^{11b_3\cdots b_m}|0\rangle$ and
$B_{1}(v_1)B_{2}(v_2)\Phi_{m-2}^{b_3\cdots
b_m}(v_3,\cdots,v_m)F^{12b_3\cdots b_m}|0\rangle$  are thought to
be linearly independent. Then, by using this assumption, the
property of Eq.(\ref{npsp}) and some necessary relations, we can
prove the  conclusions Eq.(\ref{d1pns}) as done in Ref.\cite{ik}.

In order to obtain the eigenvalue and the corresponding Bethe
equations, we need to change Eq.(\ref{d1pns}) by the following
four useful relations (the proofs are omitted here) ,
\begin{eqnarray}
& & \hspace{-6mm}S^{d_1\cdots d_m}_{c_1\cdots
c_m}(v_i;\{\check{v}_i\})\tau_1(\tilde{v}_i;\{\tilde{v}_m\})^{c_1\cdots
c_m}_{b_1\cdots b_m}=\nonumber \\
& & ({\rho^{\frac{1}{2}}_{n-1}(0)})^{-1}T^{(d_1)}(v_i)
[T^{m-1}(v_i;\{\check{v}_i\})^{d_2\cdots d_m}_{c_2'\cdots
c_m'}]_{d_1c_1'}S^{c_1'\cdots c_m'}_{b_1\cdots
b_m}(v_i;\{\check{v}_i\}), \label{ir1}
\end{eqnarray}

\begin{eqnarray}
& & \hspace{-6mm}S^{e_2\cdots e_m}_{c_2\cdots
c_m}(v_j;\{\check{v}_i,\check{v}_j\})S^{c_1\cdots c_m}_{d_1\cdots
d_m}(v_i;\{\check{v}_i\})\tau_1(\tilde{v}_i;\{\tilde{v}_m\})^{d_1\cdots
d_m}_{b_1\cdots b_m}=\nonumber \\
& &
(\rho_{n-1}(\tilde{v}_i-\tilde{v}_j){\rho^{\frac{1}{2}}_{n-1}(0)})^{-1}
T^{(c_1)}(v_i)R^{(n-1)}(\tilde{v}_i+\tilde{v}_j)^{c_1e_2}_{h_1f_1}R^{(n-1)}(\tilde{v}_i-\tilde{v}_j)^{f_1h_1'}_{c_2'd_1'}
\nonumber\\
& &\times [T^{m-2}(v_i;\{\check{v}_i,\check{v}_j\})^{e_3\cdots
e_m}_{c_3'\cdots c_m'}]_{h_1h_1'}S^{c_2'\cdots c_m'}_{d_2'\cdots
d_m'}(v_j;\{\check{v}_i,\check{v}_j\})S^{d_1'\cdots
d_m'}_{b_1\cdots b_m}(v_i;\{\check{v}_i\}), \label{ir2}
\end{eqnarray}
\begin{eqnarray}
& & \hspace{-6mm}S^{e_2\cdots e_m}_{c_2\cdots
c_m}(v_j;\{\check{v}_i,\check{v}_j\})S^{c_1\cdots c_m}_{d_1\cdots
d_m}(v_i;\{\check{v}_i\})\tau_1(\tilde{v}_j;\{\tilde{v}_m\})^{d_1\cdots
d_m}_{b_1\cdots b_m}=\nonumber \\
& &
(\rho_{n-1}(\tilde{v}_j-\tilde{v}_i){\rho^{\frac{1}{2}}_{n-1}(0)})^{-1}
T^{(h_1)}(v_j)R^{(n-1)}(\tilde{v}_j-\tilde{v}_i)^{c_1e_2}_{g_1h_1}R^{(n-1)}(\tilde{v}_j+\tilde{v}_i)^{h_1g_1}_{h_2d_1'}
\nonumber\\
& &\times [T^{m-2}(v_j;\{\check{v}_i,\check{v}_j\})^{e_3\cdots
e_m}_{c_3'\cdots c_m'}]_{h_2c_2'}S^{c_2'\cdots c_m'}_{d_2'\cdots
d_m'}(v_j;\{\check{v}_i,\check{v}_j\})S^{d_1'\cdots
d_m'}_{b_1\cdots b_m}(v_i;\{\check{v}_i\}),\label{ir3}
\end{eqnarray}
\begin{eqnarray}
& & \hspace{-6mm}S^{e_2\cdots e_m}_{a_2\cdots
a_m}(v_j;\{\check{v}_i,\check{v}_j\})S^{a_1\cdots a_m}_{c_1\cdots
c_m}(v_i;\{\check{v}_i\})\tau_1(\tilde{v}_i;\{\tilde{v}_m\})^{c_1\cdots
c_m}_{d_1\cdots d_m}\tau_1(\tilde{v}_j;\{\tilde{v}_m\})^{d_1\cdots
d_m}_{b_1\cdots b_m}\nonumber \\
& & =
(\rho_{n-1}(\tilde{v}_i-\tilde{v}_j){\rho_{n-1}(0)})^{-1}T^{(a_1)}(v_i)
T^{(f_1)}(v_j)R^{(n-1)}(\tilde{v}_i+\tilde{v}_j)^{a_1e_2}_{h_1f_1}R^{(n-1)}(\tilde{v}_j+\tilde{v}_i)^{f_1h_1'}_{f_2d_1'}
\nonumber\\
& &\times [T^{m-2}(v_i;\{\check{v}_i,\check{v}_j\})^{e_3\cdots
e_m}_{a_3'\cdots
a_m'}]_{h_1h_1'}[T^{m-2}(v_j;\{\check{v}_i,\check{v}_j\})^{a_3'\cdots
a_m'}_{c_3'\cdots c_m'}]_{f_2c_2'}\nonumber\\
& &\times S^{c_2'\cdots c_m'}_{d_2'\cdots
d_m'}(v_j;\{\check{v}_i,\check{v}_j\})S^{d_1'\cdots
d_m'}_{b_1\cdots b_m}(v_i;\{\check{v}_i\}),\label{ir4}
\end{eqnarray}
where
\begin{eqnarray}
& &[T^{m}(u;\{{v}_m\})^{d_1\cdots d_m}_{c_1\cdots
c_m}]_{ab}\nonumber \\
&&={L}(\tilde{u},\tilde{v}_1)^{ad_1}_{h_1g_1}
   {L}(\tilde{u},\tilde{v}_2)^{h_1d_2}_{h_2g_2}\cdots
   {L}(\tilde{u},\tilde{v}_m)^{h_{m-1}d_m}_{h_mg_m}k^-(u)_{h_m}\nonumber\\
& & \hspace{2mm}\times
{L^{-1}}(-\tilde{u},\tilde{v}_m)^{h_mg_m}_{f_{m-1}c_m}
{L^{-1}}(-\tilde{u},\tilde{v}_{m-1})^{f_{m-1}g_{m-1}}_{f_{m-2}c_{m-1}}
\cdots {L^{-1}}(-\tilde{u},\tilde{v}_1)^{f_1g_1}_{bc_1},
\label{trans2}
\end{eqnarray}
\begin{eqnarray}
&& T^{(d_1)}(v_i)=k^+_d(v_i)R^{(n-1)}(2v_i-4\eta)^{dd_1}_{d_1d}.
\end{eqnarray}
Let
\begin{eqnarray}
&&\Lambda_2^{m}(u; \{{v}_m\})=\prod
_{i=1}^m \rho_{n-1}(u-v_i)\tilde{\rho}(u,v_i),\\
&&\rho_{n-1}(u)=a_{n-1}(u)a_{n-1}(-u),\hspace{4mm}
 \tilde{\rho}(u,v)=\frac{1}{a_n(u+v)e_n(u-v)},
\end{eqnarray}
\begin{eqnarray}
& &
\hspace{-6mm}|\tilde{\Psi}_{m-1}^{(2)}(u,v_i;\{v_m\})_{fd}\rangle=
\frac{\rho^{\frac{1}{2}}_{n-1}(0)\omega(v_i)}{T^{(d)}(v_i)}
\Lambda_2^{m-1}(v_i;\{\check{v}_i\}) B_f(u)\Phi_{m-1}^{e_2\cdots
e_m}(v_1,\cdots,\check{v}_i,\cdots,v_m)\nonumber\\
& & \hspace{10mm} \times S^{de_2\cdots e_m}_{d_1\cdots
d_m}(v_i;\{\check{v}_i\})
\tau_1(\tilde{v}_i;\{\tilde{v}_m\})^{d_1\cdots d_m}_{b_1\cdots
b_m} F^{b_1\cdots b_m}|0\rangle,\label{tphi2}\\[4mm]
& & \hspace{-6mm}
|\tilde{\Psi}_{m-1}^{(4)}(u,v_i;\{v_m\})_{ab}\rangle=\frac{\rho^{\frac{1}{2}}_{n-1}(0)
\omega(v_i)}{T^{(\bar{b})}(v_i)}
\Lambda_2^{m-1}(v_i;\{\check{v}_i\}) E_a(u)\Phi_{n-1}^{e_2\cdots
e_m}(v_1,\cdots,\check{v}_i,\cdots,v_m)\nonumber\\
& & \hspace{10mm}\times S^{\bar{b}e_2\cdots e_m}_{d_1\cdots
d_m}(v_i;\{\check{v}_i\})
\tau_1(\tilde{v}_i;\{\tilde{v}_m\})^{d_1\cdots d_m}_{b_1\cdots
b_m} F^{b_1\cdots b_m}|0\rangle,\label{tphi4}\\[4mm]
& &\hspace{-10mm}
|\tilde{\Psi}_{m-2}^{(5)}(u,v_i,v_j;\{v_m\})_{d_1}\rangle=
F(u)\Phi_{m-2}^{e_3\cdots e_m}(v_1,\cdots,\check{v}_i,\cdots,\check{v}_j,\cdots,v_m)\nonumber \\
& &\times S^{\bar{d}_1e_3\cdots e_m}_{d_2\cdots
d_m}(v_j;\{\check{v}_i,\check{v}_j\})S^{d_1\cdots d_m}_{b_1\cdots
b_m}(v_i;\{\check{v}_i\})\nonumber \\
& & \hspace{4mm}\times \Lambda_1^{m-1}(v_i;
\{\check{v}_i\})\Lambda_1^{m-1}(v_j;
\{\check{v}_j\})\omega_1(v_i)\omega_1(v_j)F^{b_1\cdots
b_m}|0\rangle,\label{tphi5}\\[4mm]
& &\hspace{-10mm}
|\tilde{\Psi}_{m-2}^{(6)}(u,v_i,v_j;\{v_m\})_{d_1}\rangle=
F(u)\Phi_{m-2}^{e_3\cdots e_m}(v_1,\cdots,\check{v}_i,\cdots,\check{v}_j,\cdots,v_m)\nonumber \\
& &\times S^{\bar{d}_1e_3\cdots e_m}_{d_2\cdots
d_m}(v_j;\{\check{v}_i,\check{v}_j\})S^{d_1\cdots d_m}_{c_1\cdots
c_m}(v_i;\{\check{v}_i\})\tau_1(\tilde{v}_i;\{\tilde{v}_m\})^{c_1\cdots
c_m}_{b_1\cdots b_m}\nonumber \\
& & \hspace{4mm}\times \Lambda_2^{m-1}(v_i;
\{\check{v}_i\})\omega(v_i){\rho^{\frac{1}{2}}_{n-1}(0)}\Lambda_1^{m-1}(v_j;
\{\check{v}_j\})\omega_1(v_j)F^{b_1\cdots
b_m}|0\rangle,\label{tphi6}\\[4mm]
& &\hspace{-10mm}
|\tilde{\Psi}_{m-2}^{(7)}(u,v_i,v_j;\{v_m\})_{d_1}\rangle=
F(u)\Phi_{m-2}^{e_3\cdots e_m}(v_1,\cdots,\check{v}_i,\cdots,\check{v}_j,\cdots,v_m)\nonumber \\
& &\times S^{\bar{d}_1e_3\cdots e_m}_{d_2\cdots
d_m}(v_j;\{\check{v}_i,\check{v}_j\})S^{d_1\cdots d_m}_{c_1\cdots
c_m}(v_i;\{\check{v}_i\})\tau_1(\tilde{v}_j;\{\tilde{v}_m\})^{c_1\cdots
c_m}_{b_1\cdots b_m}\nonumber \\
& & \hspace{4mm}\times \Lambda_2^{m-1}(v_j;
\{\check{v}_j\})\omega(v_j){\rho^{\frac{1}{2}}_{n-1}(0)}\Lambda_1^{m-1}(v_i;
\{\check{v}_i\})\omega_1(v_i)F^{b_1\cdots
b_m}|0\rangle,\label{tphi7}\\[4mm]
& &\hspace{-10mm}
|\tilde{\Psi}_{m-2}^{(8)}(u,v_i,v_j;\{v_m\})_{d_1}\rangle=
F(u)\Phi_{m-2}^{e_3\cdots e_m}(v_1,\cdots,\check{v}_i,\cdots,\check{v}_j,\cdots,v_m)\nonumber \\
& &\times S^{\bar{d}_1e_3\cdots e_m}_{d_2\cdots
d_m}(v_j;\{\check{v}_i,\check{v}_j\})S^{{d}_1\cdots
d_m}_{c_1\cdots
c_m}(v_i;\{\check{v}_i\})\tau_1(\tilde{v}_i;\{\tilde{v}_m\})^{c_1\cdots
c_m}_{a_1\cdots a_m}\nonumber \\
& & \times \tau_1(\tilde{v}_j;\{\tilde{v}_m\})^{a_1\cdots
a_m}_{b_1\cdots b_m}\Lambda_2^{m-1}(v_i;
\{\check{v}_i\})\Lambda_2^{m-1}(v_j;
\{\check{v}_j\})\omega(v_i)\omega(v_j){\rho_{n-1}(0)}F^{b_1\cdots
b_m}|0\rangle. \label{tphi8}
\end{eqnarray}
Using relation Eq.(\ref{ir1}), we can easily change the
$|{\Psi}_{m-1}^{(2)}(u,v_i;\{v_m\})_{fd}\rangle$ and
$|{\Psi}_{m-1}^{(4)}(u,v_i;\{v_m\})_{ab}\rangle$ in
Eq.(\ref{d1pns}) into
$|\tilde{\Psi}_{m-1}^{(2)}(u,v_i;\{v_m\})_{fd}\rangle$ and
$|\tilde{\Psi}_{m-1}^{(4)}(u,v_i;\{v_m\})_{ab}\rangle$,
respectively.

 Let $e_2=\bar{c}_1$, from Eq.(\ref{ir2}), we can get
\begin{eqnarray}
& &\hspace{-6mm}
R^{(n-1)}(\tilde{v}_i-\tilde{v}_j)^{\bar{a}_1h_1'}_{c_2'd_1'}[{T}^{m-2}(v_i;\{\check{v}_i,\check{v}_j\})^{e_3\cdots
e_m}_{c_3'\cdots c_m'}]_{a_1h_1'} S^{c_2'\cdots c_m'}_{d_2'\cdots
d_m'}(v_j;\{\check{v}_i,\check{v}_j\})S^{d_1'\cdots
d_m'}_{b_1\cdots b_m}(v_i;\{\check{v}_i\})\nonumber \\
&
&\hspace{3mm}=\frac{\rho_{n-1}(\tilde{v}_i-\tilde{v}_j){\rho^{\frac{1}{2}}_{n-1}(0)}}
{\rho_{n-1}(\tilde{v}_i+\tilde{v}_j)}
\frac{R^{(n-1)}(-\tilde{v}_i-\tilde{v}_j)^{\bar{a}_1a_1}_{\bar{d}d}}{T^{(d)}(v_i)}\nonumber \\
& &\hspace{6mm}\times S^{\bar{d}e_3\cdots e_m}_{h_2\cdots
h_m}(v_j;\{\check{v}_i,\check{v}_j\})S^{{d}h_2\cdots
h_m}_{d_1\cdots
d_m}(v_i;\{\check{v}_i\})]\tau_1(\tilde{v}_i;\{\tilde{v}_m\})^{d_1\cdots
d_m}_{b_1\cdots b_m}.\label{fu22}
\end{eqnarray}

\noindent One  should note that  the repeated indices $c_1$ and
$a_1$ in Eqs.(\ref{fu22}, \ref{fu33}, \ref{fu44}) do not sum.
Using Eq.(\ref{fu22}) and  Eq.(\ref{haa2}), we have
\begin{eqnarray}
& & \hspace{-10mm}
H^{x}_{2,d_1}(u,v_i,v_j)^{fe}_{c_2d_1}|\Psi_{m-2}^{(6)}(u,v_i,v_j;\{v_m\})_{d_1c_2}^{ef}\rangle
=\nonumber \\
& &
\tilde{H}^{x}_{2,d_1}(u,v_i,v_j)|\tilde{\Psi}_{m-2}^{(6)}(u,v_i,v_j;\{v_m\})_{d_1}\rangle,
\end{eqnarray}
where
\begin{eqnarray}
& &\hspace{-10mm} \tilde{H}^{x}_{2,d_1}(u,v_i,v_j)=\frac{1}
{\tilde{\rho}(v_i,v_j)a^1_1(v_j,v_i)}
\frac{H^{x}_{2,{d}}(u,v_i,v_j)^{dd}_{dd}}
{R^{(n-1)}(\tilde{v}_i-\tilde{v}_j)^{dd}_{dd}}
\frac{R^{(n-1)}(-\tilde{v}_i-\tilde{v}_j)^{d\bar{d}}_{\bar{d}_1d_1}}{\rho_{n-1}(\tilde{v}_i+\tilde{v}_j)T^{(d_1)}(v_i)}.
\end{eqnarray}

\noindent Let $e_2=\bar{c}_1$, from Eq.(\ref{ir3}), we obtain

\begin{eqnarray}
& &\hspace{-8mm}
[{T}^{m-2}(v_j;\{\check{v}_i,\check{v}_j\})^{e_3\cdots
e_m}_{c_3'\cdots c_m'}]_{\bar{c}_1c_2'}S^{c_2'\cdots
c_m'}_{d_2'\cdots
d_m'}(v_j;\{\check{v}_i,\check{v}_j\})S^{{c}_1d_2'\cdots
d_m'}_{b_1\cdots
b_m}(v_i;\{\check{v}_i\})\nonumber \\
& & \hspace{-2mm} =\frac{{\rho^{\frac{1}{2}}_{
n-1}(0)}}{\rho_{n-1}(\tilde{v}_j+\tilde{v}_i)}
\frac{R^{(n-1)}(-\tilde{v}_i-\tilde{v}_j)^{c_1\bar{c}_1}_{e_2\bar{e}_2}
R^{(n-1)}(\tilde{v}_i-\tilde{v}_j)^{\bar{e}_2e_2}_{\bar{d}d}}
{T^{(\bar{e}_2)}(v_j)}\nonumber \\
& &\times S^{\bar{d}e_3\cdots e_m}_{h_2\cdots
h_m}(v_j;\{\check{v}_i,\check{v}_j\})S^{{d}h_2\cdots
h_m}_{d_1\cdots d_m}(v_i;\{\check{v}_i\}) \}
\tau_1(\tilde{v}_j;\{\tilde{v}_m\})^{d_1\cdots d_m}_{b_1\cdots
b_m}.\label{fu33}
\end{eqnarray}
 Then, by Eq.(\ref{fu33}), we have
\begin{eqnarray}
& & \hspace{-10mm} H^{x}_{3,d_1}(u,v_i,v_j)
|\Psi_{m-2}^{(7)}(u,v_i,v_j;\{v_m\})_{d_1d_1}\rangle
=\nonumber \\
& &
\tilde{H}^{x}_{3,d_1}(u,v_i,v_j)|\tilde{\Psi}_{m-2}^{(7)}(u,v_i,v_j;\{v_m\})_{d_1}\rangle,
\end{eqnarray}
where

\begin{eqnarray}
& &\hspace{-10mm} \tilde{H}^{x}_{3,d_1}(u,v_i,v_j)=
\frac{H^{x}_{3,{d}}(u,v_i,v_j)R^{(n-1)}(-\tilde{v}_i-\tilde{v}_j)^{d\bar{d}}_{e\bar{e}}
R^{(n-1)}(\tilde{v}_i-\tilde{v}_j)^{\bar{e}e}_{\bar{d}_1d_1}}
{\tilde{\rho}(v_j,v_i)a^1_1(v_i,v_j)\rho_{n-1}(\tilde{v}_j-\tilde{v}_i)
\rho_{n-1}(\tilde{v}_j+\tilde{v}_i)T^{(\bar{e})}(v_j)}.
\end{eqnarray}

\noindent Let $e_2=\bar{a}_1$, from Eq.(\ref{ir4}), we achieve
\begin{eqnarray}
& &\hspace{-8mm}
R^{(n-1)}(\tilde{v}_j+\tilde{v}_i)^{\bar{a}_1h_1'}_{f_2d_1'}[T^{n-2}(v_i;\{\check{v}_i,\check{v}_j\})^{e_3\cdots
e_m}_{a_3'\cdots
a_m'}]_{a_1h_1'}[T^{m-2}(v_j;\{\check{v}_i,\check{v}_j\})^{a_3'\cdots
a_m'}_{c_3'\cdots c_m'}]_{f_2c_2'}\nonumber \\
& & \hspace{-2mm}\times S^{c_2'\cdots c_m'}_{d_2'\cdots
d_m'}(v_j;\{\check{v}_i,\check{v}_j\})S^{d_1'\cdots
d_m'}_{b_1\cdots
b_m}(v_i;\{\check{v}_i\})=\frac{\rho_{n-1}(\tilde{v}_i-\tilde{v}_j){\rho_{n-1}(0)}}{\rho_{n-1}(\tilde{v}_i+\tilde{v}_j)}
\nonumber \\
& &\times
\frac{R^{(n-1)}(-\tilde{v}_j-\tilde{v}_i)^{\bar{a}_1a_1}_{\bar{d}d}}{{T^{({d})}(v_i)}{T^{(\bar{a}_1)}(v_j)}}
S^{\bar{d}e_3\cdots e_m}_{a_2\cdots
a_m}(v_j;\{\check{v}_i,\check{v}_j\})S^{{d}a_2\cdots
a_m}_{g_1\cdots g_m}(v_i;\{\check{v}_i\}) ]\nonumber\\
&&\times \tau_1(\tilde{v}_i;\{\tilde{v}_m\})^{g_1\cdots
g_m}_{d_1\cdots d_m}\tau_1(\tilde{v}_j;\{\tilde{v}_m\})^{d_1\cdots
d_m}_{b_1\cdots b_m}.\label{fu44}
\end{eqnarray}

\noindent Using Eq.(\ref{fu44}) and considering Eq.(\ref{haa4}),
we have
\begin{eqnarray}
& & \hspace{-10mm}
H^{x}_{4,d_1}(u,v_i,v_j)^{de}_{fd_1}|\Psi_{m-2}^{(8)}(u,v_i,v_j;\{v_m\})_{d_1f}^{ed}\rangle
=\nonumber \\
& &
\tilde{H}^{x}_{4,d_1}(u,v_i,v_j)|\tilde{\Psi}_{m-2}^{(8)}(u,v_i,v_j;\{v_m\})_{d_1}\rangle,
\end{eqnarray}
where

\begin{eqnarray}
& &\hspace{-6mm} \tilde{H}^{x}_{4,d_1}(u,v_i,v_j)=\frac{1}
{\tilde{\rho}(v_i,v_j)\tilde{\rho}(v_j,v_i)\rho_{n-1}(\tilde{v}_j-\tilde{v}_i)\rho_{n-1}(\tilde{v}_i+\tilde{v}_j)}\nonumber\\
& & \hspace{25mm}\times\frac{H^{x}_{4,{d}}(u,v_i,v_j)^{dd}_{dd}}
{R^{(n-1)}(\tilde{v}_i-\tilde{v}_j)^{dd}_{dd}}
\frac{R^{(n-1)}(-\tilde{v}_j-\tilde{v}_i)^{d\bar{d}}_{\bar{d}_1d_1}}{{T^{({d_1})}(v_i)}{T^{(d)}(v_j)}}.
\end{eqnarray}
We can easily get
\begin{eqnarray}
& & \hspace{-10mm} \delta_{\bar{d}_1e_2}
H^{x}_{1,d_1}(u,v_i,v_j)|\Psi_{m-2}^{(5)}(u,v_i,v_j;\{v_m\})_{d_1e_2}\rangle
=\nonumber \\
& &
\tilde{H}^{x}_{1,d_1}(u,v_i,v_j)|\tilde{\Psi}_{m-2}^{(5)}(u,v_i,v_j;\{v_m\})_{d_1}\rangle,
\end{eqnarray}
where

\begin{eqnarray}
& &\hspace{-10mm} \tilde{H}^{x}_{1,d_1}(u,v_i,v_j)=
\frac{H^{x}_{1,d_1}(u,v_i,v_j)}{a^1_1(v_i,v_j)a^1_1(v_j,v_i)}.
\end{eqnarray}
After making the  notation
\begin{eqnarray}
|\tilde{\Psi}_{x}(u,\{v_m\})\rangle=\left\{\begin{array}{l}|{\Psi}_{x}(u,\{v_m\})\rangle, x=A,\tilde{A}_2\\[2mm]
\omega(u)\Lambda_2^m(u;\{{v}_m\})\Phi_m^{d_1\cdots
d_m}(v_1,\cdots,v_m)[{T}^{m}(u;\{{v}_m\})^{d_1\cdots
d_m}_{b_1\cdots b_m}]_{aa} F^{b_1\cdots b_m}|0\rangle,
x=\tilde{A}_{aa} \end{array}\right.
\end{eqnarray}
we arrive at the final result Eq.(\ref{nd1pns}).



\begin{thebibliography}{25}



\bibitem{fad} L.D. Faddeev,  1982 {\it In recent advances in Field theory and Statistical
Mechanics}, Les Houches.

\bibitem{7}  V.E.Korepin,  N.M.Bogoliubov and  A.G.Izergin   1993 {\it Quantum Inverse
Scattering Method and Correlation Function}, Cambridge University
Press. Cambridge.
\bibitem{8}  E. K.Sklyanin, 1988 { \sl J. Phys.} {\bf  A 21}
2375.

\bibitem{alcar} F. C. Alcaraz, M. N. Barber, M. T. Batchelor, R.
J. Baxter and G. R. W. Quispel, 1987 { \sl J. Phys.} {\bf  A 20}
6397.



\bibitem{an} O. Babelon, H.J. De Vega and C.M. Viallet,  1982 {\sl Nucl.
Phys.} {\bf B 200} (1982) 266.
\bibitem{ano} H.J.de vega and A. Gonzalez-Ruiz, 1994 {\sl Nucl. Phys.} {\bf B 417}
 553 [hep-th/9404141].


\bibitem{9}  P.B.Ramos and M.J.Martins,
1996 {\sl Nucl. Phys.} {\bf B 474} 678 [hep-th/9604072].
\bibitem{10}  M.J.Martins and  P.B.Ramos,  1997 { \sl Nucl. Phys.} {\bf B 500}  579 [hep-th/9703023].
\bibitem{11} M.J.Martins and P.B.Ramos, 1998
{\sl Nucl. Phys.} {\bf B 522}  413 [solv-int/9712014].
\bibitem{112} M.J.Martins, 1999 {\sl Phys.Rev.} {\bf E 59}  7220 [solv-int/9901002].
\bibitem{12} W. Galleas and M.J.Martins, 2004
{\sl Nucl. Phys.} {\bf B 699}  455.

\bibitem{13} X.W. Guan, 2000 {\sl J. Phys.} {\bf A 33}  5391.

\bibitem{14} A.Foerster,  X.W.Guan,  J.Links,  I.Roditi and
H.Q.Zhou, 2001 {\sl Nucl. Phys.} {\bf B 596}  525.

\bibitem{15}  X.W.Guan, A.Foerster, U.Grimm, R.A.R{\"{o}}mer and
M.Schreiber, 2001 {\sl Nucl. Phys.} {\bf B 618}  650.

\bibitem{ik}  G.L.Li,  K.J.Shi and  R.H.Yue,  2003 {\sl Nucl.Phys.} {\bf B 670}
401.
\bibitem{osp} G.L.Li,  K.J.Shi and  R.H.Yue, 2004 {\sl Nucl. Phys. } {\bf B 687}
(2004) 220.\\
 G.L.Li,  K.J.Shi and  R.H.Yue, 2005 {\sl J. High
Energy Phys.} {\bf 0507}  001 [hep-th/0505001].
\bibitem{osp12} V. Kurak, A. Lima-Santos, 2004 {\sl Nucl.Phys.} {\bf B
699}  595 [nlin.SI/0406050].\\
V. Kurak, A. Lima-Santos, 2005 {\sl J. Phys.} {\bf A 38} 2359
[nlin.SI/0407006].


\bibitem{1}  N.Yu.Reshetikhin, 1987
{\sl Lett. Math.Phys.} {\bf 14}  235.

\bibitem{2}  L.Mezincescu and R.I. Nepomechie, 1992 {\sl Nucl. Phys. } {\bf B 372}  597 [hep-th/9110050].

\bibitem{a2n2}  S.Artz, L.Mezincescu and R.I. Nepomechie, 1995 {\sl Int. J.
Mod.Phys.} {\bf A 10}  1937 [hep-th/9409130].\\
S.Artz, L.Mezincescu and R.I. Nepomechie, 1995 {\sl J. Phys.} {\bf
A 28} 5131 [hep-th/9504085].



\bibitem{lima} A. Lima-santos, 2006
{\sl J.Stat.Mech}   P07003 [nlin.SI/0602003].


\bibitem{vv}  V.V.Bazhanov, 1985
  {\sl Phys. Lett. } {\bf B 519}  321.\\
  V.V.Bazhanov, 1987 {\sl Commun. Math. Phys.} {\bf 113} 471.

\bibitem{mj} M.Jimbo, 1986 {\sl Commun. Math. Phys.} {\bf 102}
537.





\bibitem{A2n}  A.Lima-santos, 2003 {\bf B 654} 466
[nlin.SI/0210046].\\
A.Lima-santos  and  R. Malara R,  2003 {\sl Nucl.Phys.} {\bf B
675} 661 [nlin.SI/0307046].
\bibitem{A2n1} R. Malara and A.Lima-santos, J. Stat. Mech. (2006)
P09013.

\bibitem{6}  R.J.Baxter,   1982 { \it Exactly Solved Models in Statitical
Mechanics}, Academic Press, London.

\bibitem{cao} J.P.Cao, H.Q.Lin, K.J.Shi and Y.P.Wang, 2003 {\sl Nucl.Phys.} {\bf B 663} 487.

\bibitem{} W.Galleas and M.J.Martins, 2005 {\sl Phys.Lett.} {\bf A 335}
167 [nlin.SI/0407027].

\bibitem{xxx}  C.S.Melo, G.A.P.Ribeiro and M.J.Martins, 2005 {\sl Nucl. Phys. } {\bf B 711} 565.

\bibitem{yangwl} W.L. Yang and Y.Z. Zhang, 2005 {\sl J. High
Energy Phys.} {\bf 0501}  021 [hep-th/0411190].




\end{thebibliography}
\end{document}